\documentclass{aa}  
\usepackage[dvips]{graphicx}
\usepackage{psfig}
\def\N2H{N$_2$H$^+$}
\def\CSI{C$^{34}$S}
\def\CS{C$^{32}$S}

\def\COI{$^{13}$CO}
\def\COII{C$^{18}$O}

\def\HM{H$_2$}

\def\Vlsr{\mbox{$V_{\rm LSR}$}}
\def\Kkms{K~km~s$^{-1}$}
\def\kms{km~s$^{-1}$}
\def\kp{km~s$^{-1}$~pc$^{-1}$}
\def\cmc{cm$^{-3}$}

\def\nmh{\mbox{$n_{\rm H_2}$}}

\def\tmb{\mbox{$T_{\rm mb}$}}

\def\tex{\mbox{$T_{\rm ex}$}}
\def\tbg{\mbox{$T_{\rm bg}$}}
\def\tk{\mbox{$T_{\rm k}$}}

\def\vl{V_{\rm LSR}}
\def\solmass{M$_\odot$}

\def\bra{\left <}
\def\ket{\right >}
\def\la{\mathrel{\mathchoice {\vcenter{\offinterlineskip\halign{\hfil
$\displaystyle##$\hfil\cr<\cr\sim\cr}}}
{\vcenter{\offinterlineskip\halign{\hfil$\textstyle##$\hfil\cr
<\cr\sim\cr}}}
{\vcenter{\offinterlineskip\halign{\hfil$\scriptstyle##$\hfil\cr
<\cr\sim\cr}}}
{\vcenter{\offinterlineskip\halign{\hfil$\scriptscriptstyle##$\hfil\cr
<\cr\sim\cr}}}}}
\def\ga{\mathrel{\mathchoice {\vcenter{\offinterlineskip\halign{\hfil
$\displaystyle##$\hfil\cr>\cr\sim\cr}}}
{\vcenter{\offinterlineskip\halign{\hfil$\textstyle##$\hfil\cr
>\cr\sim\cr}}}
{\vcenter{\offinterlineskip\halign{\hfil$\scriptstyle##$\hfil\cr
>\cr\sim\cr}}}
{\vcenter{\offinterlineskip\halign{\hfil$\scriptscriptstyle##$\hfil\cr
>\cr\sim\cr}}}}}

\newcommand{\beq}{\begin{equation}}
\newcommand{\eeq}{\end{equation}}
\newcommand{\bdi}{\begin{displaymath}}
\newcommand{\edi}{\end{displaymath}}

\begin{document}

\title{Constraints on star formation theories from the Serpens molecular cloud
and protocluster}
\author{L. Olmi \inst{1,} \inst{2} \and L. Testi\inst{3} }
\institute{LMT/GTM Project,  Dept. of Astronomy, 815J Lederle GRT Tower B, 
          University of Massachusetts, 710 N. Pleasant st. 
          Amherst, MA 01003--9305, U.S.A., {\it olmi@lmtgtm.org} 
          \and
          {\it Present address}:
          University of Puerto Rico, Dept. of Physics, PO Box 23343
          University Station, S. Juan PR 00931-3343, USA, and
          CNR - Istituto di Radioastronomia, Largo E. Fermi 5,
          I-50125 Firenze, Italy, {\it olmi@naic.edu} 
          \and
	  Osservatorio Astrofisico di Arcetri, Largo E. Fermi 5,
          I-50125 Firenze, Italy, {\it ltesti@arcetri.astro.it}  
          }
\offprints{L. Olmi}
\date{Received date; accepted date}
%
%
\abstract{
We have mapped the large-scale structure of the Serpens cloud core
using moderately optically thick ($^{13}$CO(1--0) and
CS(2--1)) and optically thin tracers (C$^{18}$O(1--0), C$^{34}$S(2--1), and
N$_2$H$^+$(1--0)), using the 16-element focal plane array
operating at a wavelength of 3mm at the Five College Radio Astronomy
Observatory. 
Our main goal was to study the large-scale distribution of the molecular
gas in the Serpens region and 
to understand its relation with the denser
gas in the cloud cores, previously studied at high angular resolution.
All our molecular tracers show two main gas condensations, or sub-clumps, 
roughly corresponding to the North-West and South-East clusters 
of submillimeter continuum sources.
We also carried out a kinematical study of the Serpens cloud. 
The  \COI\ and \COII(1--0) maps of the centroid velocity show an increasing,
smooth gradient in velocity from East to West, which we think
may be caused by a global rotation of the Serpens molecular cloud whose 
rotation axis is roughly aligned in the SN direction.
Although it appears that
the cloud angular momentum is not sufficient for being dynamically important
in the global evolution of the cluster, the fact that the observed
molecular outflows are roughly aligned with it may suggest a link between 
the large-scale angular momentum and the circumstellar disks around 
individual protostars in the cluster. 
We also used the normalized centroid velocity difference as an infall 
indicator. We find two large regions of the map, approximately coincident 
with the SE and NW sub-clumps, which are undergoing an infalling motion.
Although our evidence is not conclusive, our data appear to be in qualitative
agreement with the expectation of a slow contraction followed by a rapid
and highly efficient star formation phase in localized high density regions.
\keywords{ISM: molecules -- stars: formation -- 
          radio lines: ISM}
}
\titlerunning{The Serpens molecular cloud and protocluster}
\maketitle


\section{Introduction}
\label{sec:intro}

It is now well established that a substantial fraction 
of the young pre-main sequence stars in molecular clouds 
are found in compact groups (clusters) rather than being uniformly 
distributed throughout the clouds (e.g.  Gomez et al.~\cite{Gea93}; 
Testi et al.~\cite{TPN99}; Carpenter~\cite{C00}).
It is thus clear that the formation and early evolution
of stellar clusters hold the key for the understanding of the galactic 
disk stellar population, such as the distribution of stellar masses at 
birth, or initial mass
function (IMF), as well as the kinematical and multiplicity properties 
(Clarke et al.~\cite{CBH00}; Kroupa~\cite{K95}; Kroupa et al.~\cite{Kea01}). 
While there is a general consensus that the latter two
properties are linked to the clusters dynamical evolution, the nature of the
IMF is still debated. Observations show that the emergent mass distribution of
young stellar clusters is consistent with the field stars IMF
(Palla \& Stahler~\cite{PS99}; Hillenbrand \& Carpenter~\cite{HC00}; Meyer et
al.~\cite{Mea00}), implying that the stellar mass distribution must be 
linked with the cluster formation process itself. Three main 
classes of models have been considered: those which link the final stellar 
masses to (i) the structure
or fragmentation of  the parent molecular cloud (e.g. Elmegreen~\cite{E97};
Myers~\cite{M00}; Padoan \& Nordlund~\cite{PN01}); or (ii) the feedback of the 
protostellar accretion process (Adams \& Fatuzzo~\cite{AF96}); or, finally,
(iii) competitive accretion between already formed (proto--)stars in 
clusters (Bonnell et al.~\cite{Bea97};~\cite{Bea01}).

The progressive improvement in millimeter wavelength techniques, in particular the
development of bolometer arrays for large single dish telescopes and the 
upgrade of millimeter wave arrays,  have provided a new wealth of 
observational data to constrain theories (Andr\'e \& Motte~\cite{AM00}; 
Testi \& Sargent~\cite{TS00}).
Millimeter continuum surveys of the cluster forming clouds in Serpens (Testi \&
Sargent~\cite{TS98}, hereafter TS98), $\rho$--Oph (Motte et al.~\cite{MAN98}; 
Johstone et al.~\cite{Jea00}), NGC~1333 (Sandell \& 
Knee~\cite{SK01}), and Orion~B (Johnstone et al.~\cite{Jea01}; Motte et
al.~\cite{Mea01}) have shown that the mass distribution of the prestellar 
cores, progenitors of  individual stellar systems, is consistent with the field 
stars IMF and thus significantly steeper that the mass function of the 
larger gaseous
clumps in molecular clouds (e.g. Williams et al.~\cite{WBM00}).  These
observations suggest that the self-similarity of molecular clouds breaks down at
the cores level, and the fragmentation or the small-scale structure of the 
clouds is the likely mechanism responsible for the stellar IMF.

In a first attempt to link the small-scale structure of the prestellar 
cores and young stellar objects in clusters with the kinematics and 
large-scale distribution of the molecular gas, Testi et al.~(\cite{tsoo00}; 
hereafter TSOO) combined the high resolution 3~mm continuum survey of TS98 
with new interferometric and single dish
molecular line surveys of the Serpens core. The high resolution CS(2--1)
observations showed high velocity molecular gas associated with outflows 
emanating from several of the protocluster members, while the single 
dish N$_2$H$^+$(1--0)
revealed the presence of two main gaseous clumps, well separated in velocity,
in which the TS98 sources are embedded.
The combination of the new high and low resolution millimeter-wave data with
previously published observations showed that the protocluster is structured
in sub-clusters, each internally coherent, well separated both spatially and
kinematically. The star formation efficiency and proto-stellar density within
each core is a factor of $\sim$10 higher than the average values.
The fact that the molecular gas within each sub-cluster is kinematically 
coherent
suggests that clusters are assembled through fragmentation of the molecular
clump in coherent sub-clumps which produce independent sub-clusters (TSOO).
It is not clear, however, whether
the sub-clusters in Serpens will eventually merge into a final cluster,
such as other young clusters (e.g. the
Orion Nebula Cluster) which show little evidence for sub-structures 
(Clarke et al.~\cite{CBH00}).

Similar results on the spatial segregation of pre-stellar cores
have also been found in the $\rho$-Oph star forming region (Johstone et
al.~\cite{Jea00}), but unfortunately the essential kinematical information
is missing in this region. In both regions, the prestellar cores and protostars
densities are too low for dynamical interactions to be efficient over the
timescale in which the bulk of the mass of the final stars is assembled.
Dynamical interaction and competitive accretion can possibly play a role
for the final tuning of stellar masses but cannot play a major role in
determining the global shape of the stellar IMF in these regions.
Note however that both these regions are at least an order of magnitude
less dense than the typical clusters associated with massive stars
(such as the Orion Nebula Cluster).

Given for granted that, at least in Serpens and $\rho$-Oph, the stellar IMF
appears to be linked to the individual small-scale cloud fragments, each 
progenitor of a single stellar system, the question remains on how the 
fragmentation and collapse processes occur. There are currently 
two opposite views of these processes: a ``slow'' and a ``fast'' mode.
In the first mode molecular clouds may contract as a whole
in a slow fashion evolving through a series of quasi-equilibrium conditions;
star formation is triggered in localized regions of the cloud when the 
density exceeds the threshold for gravitational collapse. Star formation
in the cloud as a whole may be regulated by external agents, such as in the
photoionization-regulated view of
McKee~(\cite{MK89}), or it may progressively accelerate until the feedback 
of newly formed stars disperse the molecular material, halting star formation
(Palla \& Stahler~\cite{PS00}). In the ``fast'' mode, since turbulence
is expected to decay on very short timescales within molecular clouds
(Mac~Low~\cite{McL99}), there is no force to balance the clouds against
gravitational collapse and star formation occurs
 globally on a few free-fall 
times (Elmegreen~\cite{E00}; Hartmann et al.~\cite{Hea01}). From
the observational point of view, the problem can be
approached by studying the large-scale kinematical and physical conditions 
in star forming molecular clouds and compare them with position, mass and 
age distributions of forming stars (Palla \& Stahler~\cite{PS00}).

The Serpens cloud core appears to be an ideal place to test the predictions 
of the various models. At a distance of $\sim$310~pc 
(de Lara et al.~\cite{dL91}), the
Serpens cloud core and protocluster have been the subject of numerous detailed
observational studies in recent years, at infrared and (sub-)millimeter 
wavelengths. Among the most recent are: Davis et al.~\cite{dav99};
Giovannetti et al.~\cite{Gea98}; Hodapp~\cite{H99}; Hogerheijde et 
al.~\cite{Hea99};
Kaas~\cite{K99}; Kaas \& Bontemps~\cite{KB01}; McMullin et al.~\cite{mcm94};
\cite{mcm00}; Williams \& Myers~\cite{WM99}; \cite{WM00};
in addition to TS98 and TSOO. The reasons for these numerous observational 
efforts lie in its compactness, richness and relative proximity to the Sun, 
which make it possible to obtain and combine
high resolution observations of the relatively large number
of (proto-)stellar sources within the (proto-)cluster and large-scale 
observations
in various tracers of the entire cloud core. Regions further away from the Sun
cannot be adequately resolved with single dish telescopes
and cannot be observed at the required sensitivity
with current millimeter arrays, whereas nearby 
regions require a substantially
larger observing time to be fully mapped.

Most of the previous single dish millimeter studies of the large-scale 
structure of the molecular clump, within which the (proto-)cluster is 
embedded, focused either on the study of molecular outflows by means 
of optically thick CO(2--1) observations
(Davis et al.~\cite{dav99}), or on the chemistry (McMullin et 
al.~\cite{mcm94}; \cite{mcm00}), or presented partial maps of the cloud core (White, 
Casali \& Eiroa~\cite{Wea95}).
In this paper we complement these and our previous high resolution studies
(TS98 and TSOO) with observations of the molecular clump large-scale 
structure
using complete maps in moderately optically thick ($^{13}$CO(1--0) and
CS(2--1)) and optically thin tracers (C$^{18}$O(1--0), C$^{34}$S(2--1), and
N$_2$H$^+$(1--0)), taking advantage of the 16-element focal plane array
operating at a wavelength of 3mm at the  Five College Radio Astronomy 
Observatory\footnote{The Five College Radio Astronomy Observatory is 
operated with support from the National Science Foundation and with 
permission of the Metropolitan District Commission} (FCRAO).
These new observations are used to derive refined values of the physical
parameters of the cloud, and are combined with previous observations
to study the kinematical and fragmentation properties of the cloud core.

\begin{figure}
%
%
\vspace*{4mm}
\centerline{\psfig{figure=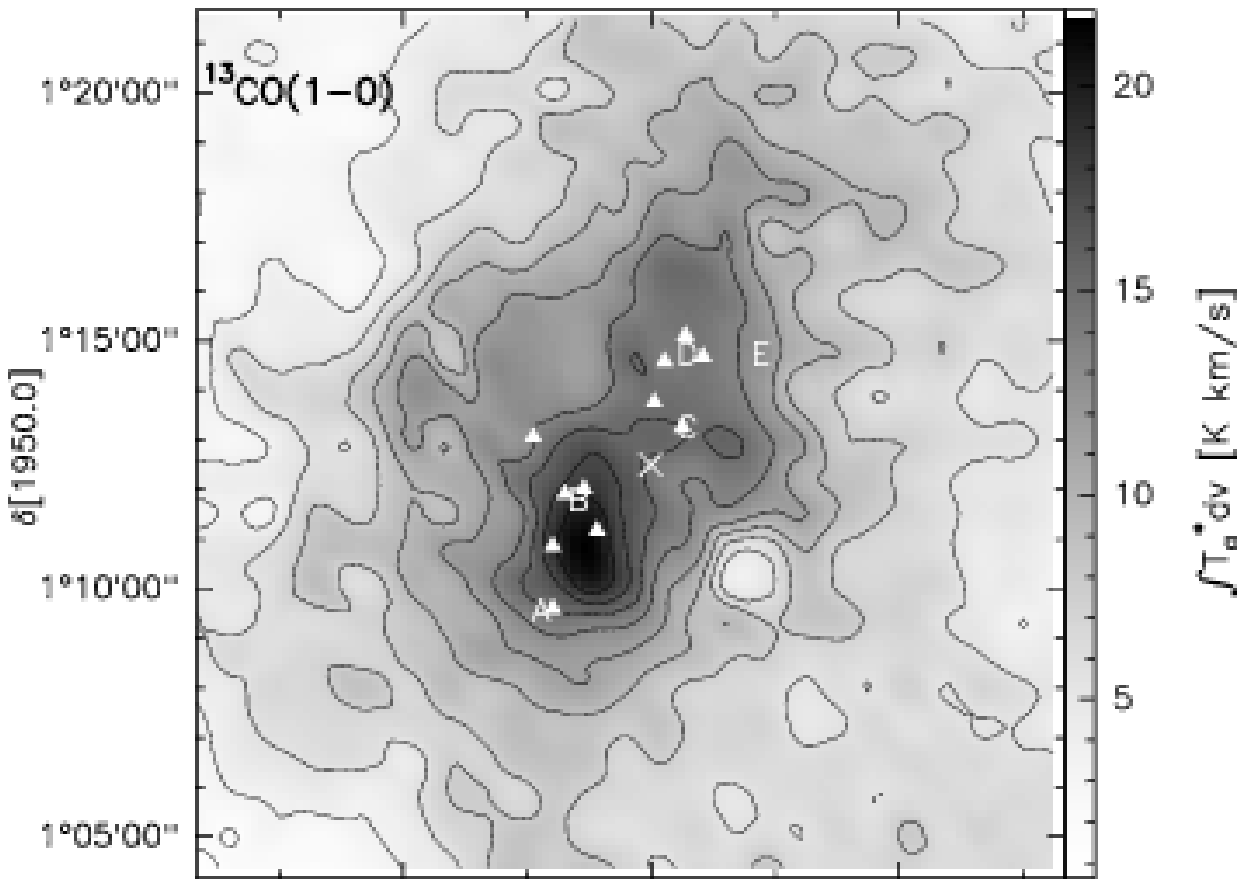,width=6.8cm,angle=270}}
\centerline{\psfig{figure=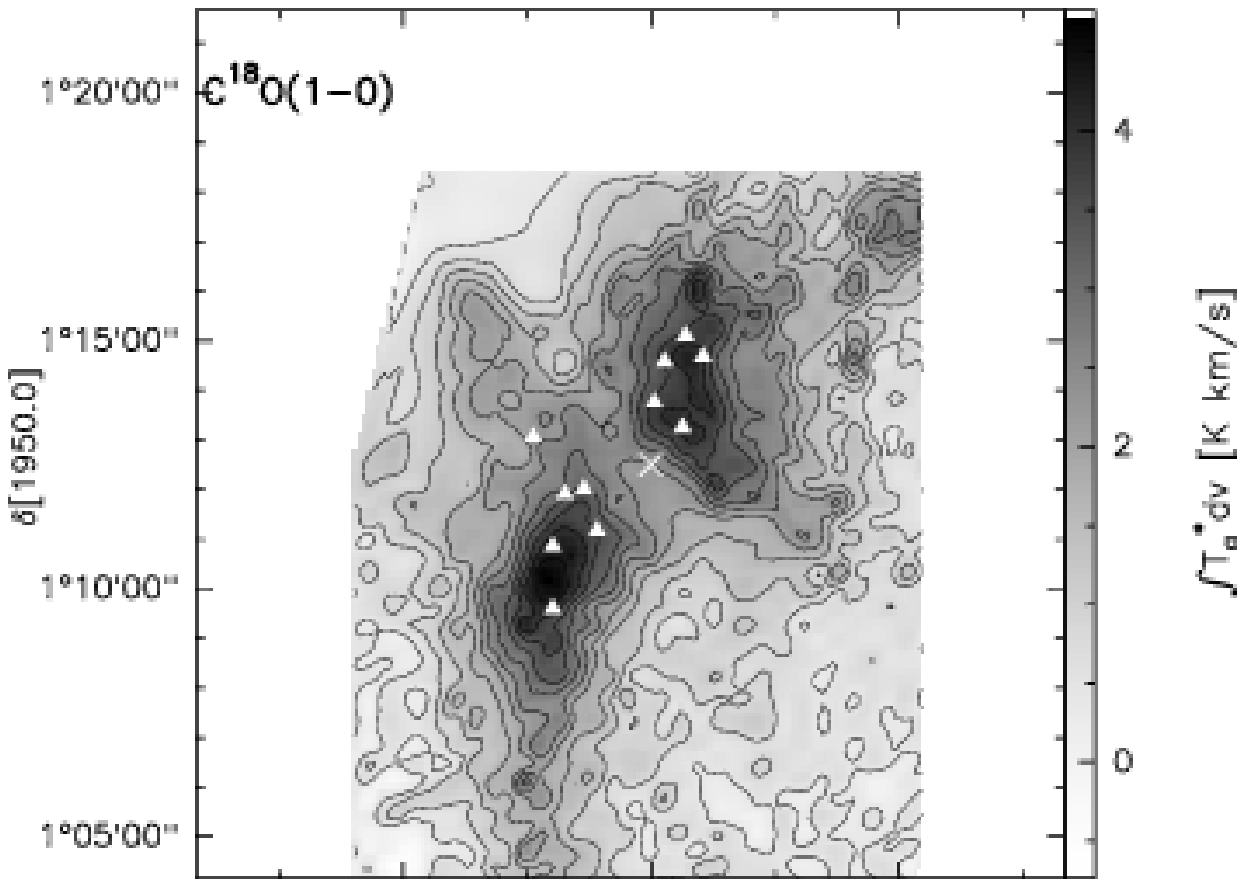,width=6.8cm,angle=270}}
\centerline{\psfig{figure=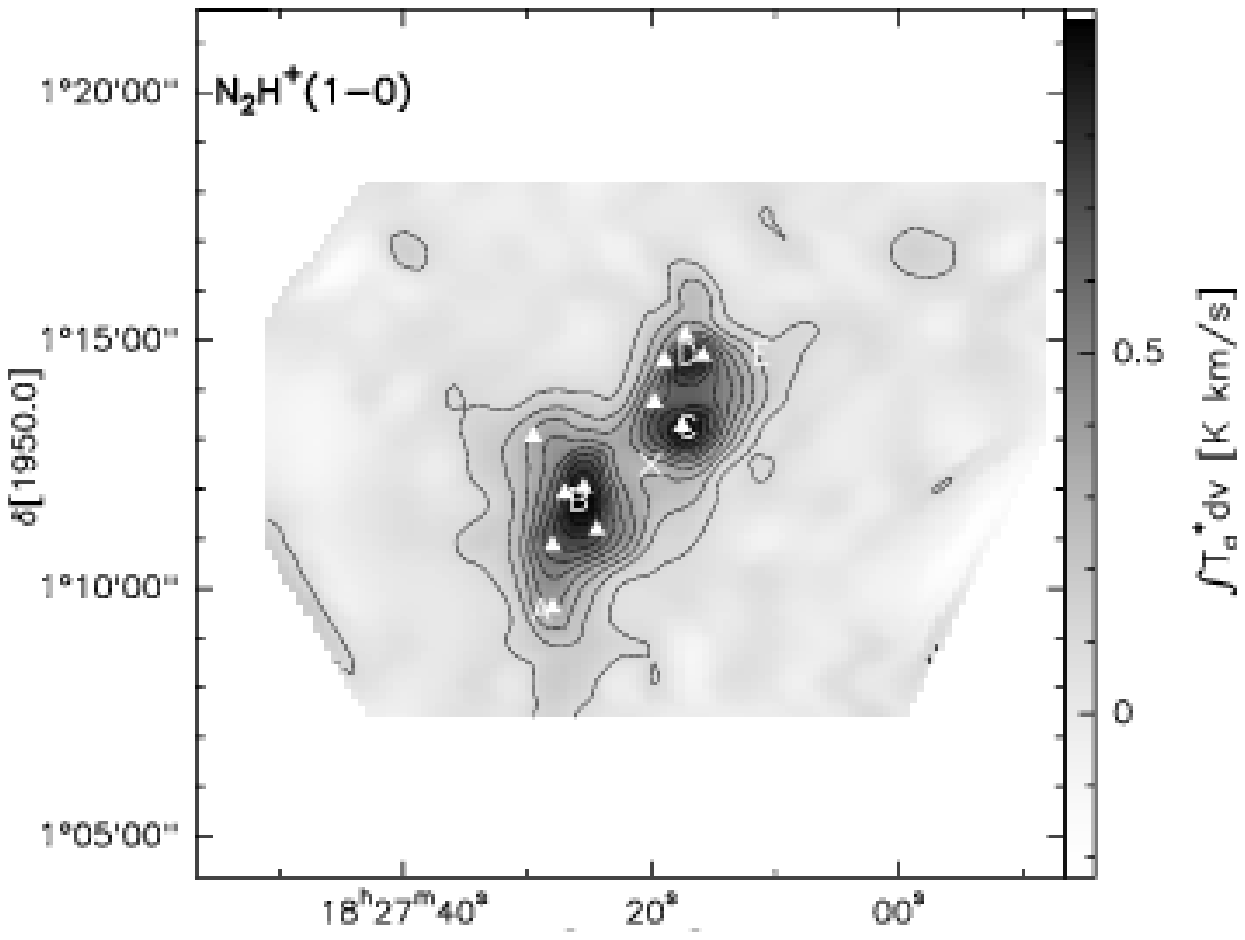,width=6.8cm,angle=270}}
\caption[ ]{
Maps of the \COI, \COII\ and \N2H\ integrated intensity in the velocity
interval 4.0 to 12.0~\kms. In the \COI\ map the value of the first
contour is 1.5~\Kkms and the contour interval is 1.0~\Kkms.
In the \COII\ and \N2H\ maps the value of the first contour and the contour
interval are both 0.4~\Kkms. 
The cross marks the 
reference position (see  Table~\ref{tab:lines}) and the filled triangles
represent the positions of the submillimeter continuum sources. 
The letters mark the positions of the main \N2H\ cores as defined in TSOO.
The ``spur'' is also visible, an extended, lower surface brightness narrow 
region, stretching from the NW to SE clumps and continuing beyond these
(see also Fig.~\ref{fig:SCUBA850}).
}
\label{fig:comap}
\end{figure}

The paper is organized as follows: in Sect.~\ref{sec:due} and~\ref{sec:mor}
we describe the new observations, the morphology of the Serpens cloud in
the observed tracers and derive the physical conditions of the molecular 
gas; in Sect.~\ref{sec:disc} we discuss the cloud kinematical and
fragmentation properties; finally, in Sect.~\ref{sec:concl} we summarize the
main results of this study, and
we discuss the implications for cluster formation and evolution theories.

\section{Observations }
\label{sec:due}

%
%
\begin{table}
\begin{flushleft}
\caption[ ]{
Molecular transitions observed and their rest frequencies. The third column
lists the size of the map centered about the position
$\alpha=18^{\rm h}27^{\rm m}20^{\rm s}$ and $\delta=1^{\circ}12'30''$  (B1950.0)
}
\begin{tabular}{lcr}
\hline
Line   & Frequency [GHz]  & Map size [arcmin] \\
\hline
CS(2--1)     &  97.980968   & $13\times10$ \\
\CSI(2--1)   &  96.412982   & $5\times5$   \\
\N2H(1$_{23}$--0$_{12}$)   & 93.1734796   & $13\times10$ \\
\COI(1--0)   & 110.201353   & $16\times16$ \\
\COII(1--0)  & 109.782182   & $12\times14$ \\
\hline
\end{tabular}
\label{tab:lines}
\end{flushleft}

\end{table}

The  single dish  observations of the  Serpens cloud core
were carried out in March, October and November 1999 with the
13.7-m telescope of the FCRAO located in New Salem (U.S.A.), using the 
new SEQUOIA 32 element focal plane array, although for the present 
observations only 16 pixels were actually used.  The observed lines and 
their rest frequencies are listed in Table~\ref{tab:lines}. 
The maps were centered about the position 
$\alpha=18^{\rm h}27^{\rm m}20^{\rm s}$ 
and $\delta=1^{\circ}12'30''$  (B1950.0), 
and the FWHM of the FCRAO 
telescope varied from about $46''$ to about $53''$.
The front-end receivers employed low noise InP MMIC-based amplifiers
resulting in a mean receiver temperature of 70~K (SSB) and
system temperatures typically in the range $160-250$~K.
Our spectrometer was an autocorrelator with 24~KHz (0.077~\kms) spectral
resolution and 20~MHz bandwidth.

The integration times used for the single-scan were typically 5 to
25 minutes in frequency switching mode, using a frequency throw of 8~MHz.
The maps were carried out using 1-beam sampling and the sizes of the 
covered regions
in the different lines are also listed in Table~\ref{tab:lines}.
The main beam efficiency used to convert the antenna temperature,
$T_{\rm A}^{*}$, to main beam brightness temperature is $\eta_{\rm mb}=0.55$.
The final achieved sensitivities in the $T_{\rm A}^{*}$ scale varied from about
0.05~K for \CSI\ to about 0.19~K for \COI.

\section{Results}
\label{sec:mor}

\subsection{Morphology of the Serpens region}
\label{sec:morph}

\subsubsection{\COI(1--0) and \COII(1--0)}
\label{sec:comorph}

We performed wide-field imaging of the \COI\ and \COII(1--0) emission 
towards the Serpens region to understand the
relation of the global distribution of the molecular gas with the denser gas 
in the cloud cores. The optically thicker \COI(1--0) line emission tends 
to pick up the more diffuse ambient gas, whereas the thinner transition
of \COII(1--0) shows the distribution of the gas inside the Serpens cloud.

The \COI(1--0), \COII(1--0) and \N2H(1--0)
 emission integrated over the velocity interval 4.0 
to 12.0~\kms, shown in Fig.~\ref{fig:comap}, trace the distribution of 
the column density in the Serpens cloud (see also Sect.\ref{sec:exc}).
The \COI\ emission is extended and encompasses both the NW and SE clusters
of submillimeter continuum sources, with a peak of emission near the 
position of SMM4. We clearly detect \COI(1--0) at the position of SMM1,
where the BIMA observations of McMullin et al.~(\cite{mcm00}) did not detect
any emission.
Compared to the \COII(1--0) map of McMullin et al.~(\cite{mcm00}) our 
\COII\ integrated intensity map better shows the clear demarcation between
the NW and SE sub-clusters, which are also separated in velocity (see 
Sect.~\ref{sec:kin} and TSOO).

\begin{figure}
%
%
\centerline{\psfig{figure=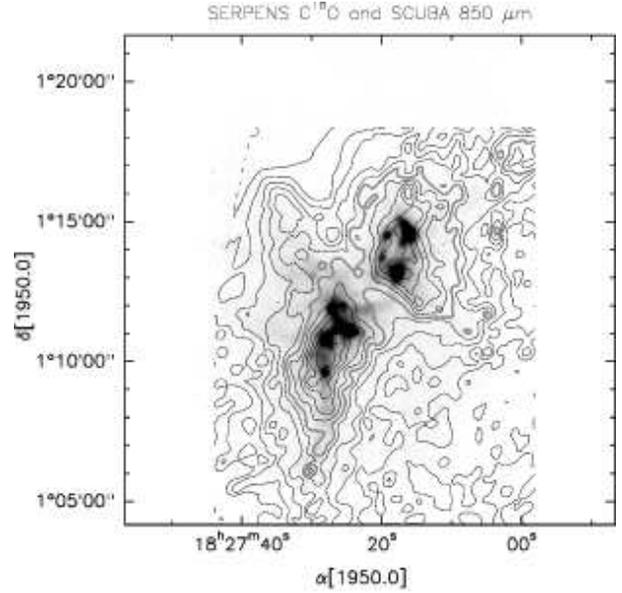,width=8.0cm,angle=270}}
\caption[ ]{
Overlay of SCUBA 850~$\mu$m gray-scale image (kindly provided by C.J. Davis, 
Davis et al.~\cite{dav99}) and
\COII(1--0) integrated emission (contour levels as in Fig.~\ref{fig:comap}).
The thicker grey contour corresponds to 50\% of the maximum.
}
\label{fig:SCUBA850}
\end{figure}

The NW clump is clearly surrounding all continuum sources: S68N 
(McMullin et al.~\cite{mcm94}) and
SMM1 in particular (also seen at the position of the \N2H\ core C) are very
close to two \COII\ emission peaks. The SE cluster of submillimeter continuum
sources lies essentially in the region between the two  \COII\ peaks, 
and is aligned with the  \COII\ lower surface brightness region extending 
to the NE.
On the other hand, sources SMM2 and SMM11 are embedded in the 
SE gas clump, and the emission is elongated in the NS direction 
along the ``spur'', 
an extended, lower surface brightness narrow region, 
stretching from the NW to SE clumps and continuing beyond these, observed
by TSOO. 
As observed by McMullin et al.~(\cite{mcm00}), in the case of
the \COII\ emission there is a lack of correspondence with the 
submillimeter peaks, especially in the SE sub-clump. However, the \COII\
and submillimeter dust continuum emission both trace the large-scale 
distribution of column density, as shown in Fig.~\ref{fig:SCUBA850}.
A similar structure to  \COII\ can be seen in \N2H (TSOO), although the peaks of
emission show a different position in the SE clump.

%
%
Fig.~\ref{fig:SCUBA850} shows that the Spur emission has a striking 
correspondence with the 850~$\mu$m continuum emission of Davis et 
al.~(\cite{dav99}) extending southward from SMM11. 
%
%
Another dust lane is extending roughly
NE from SMM3 and coincides quite well with a ``bulge'' of higher velocity
gas visible in both the \COI\ and \COII\ centroid velocity maps 
(Fig.~\ref{fig:13cocv}).
Here, as suggested by Davis et al.~(\cite{dav99}), the high-velocity gas and  
dust continuum emission are maybe showing the action of a wind powered by 
the SMM 2/3/4 group.

\begin{figure}
%
%
%
%
\vspace*{4mm}
\centerline{\psfig{figure=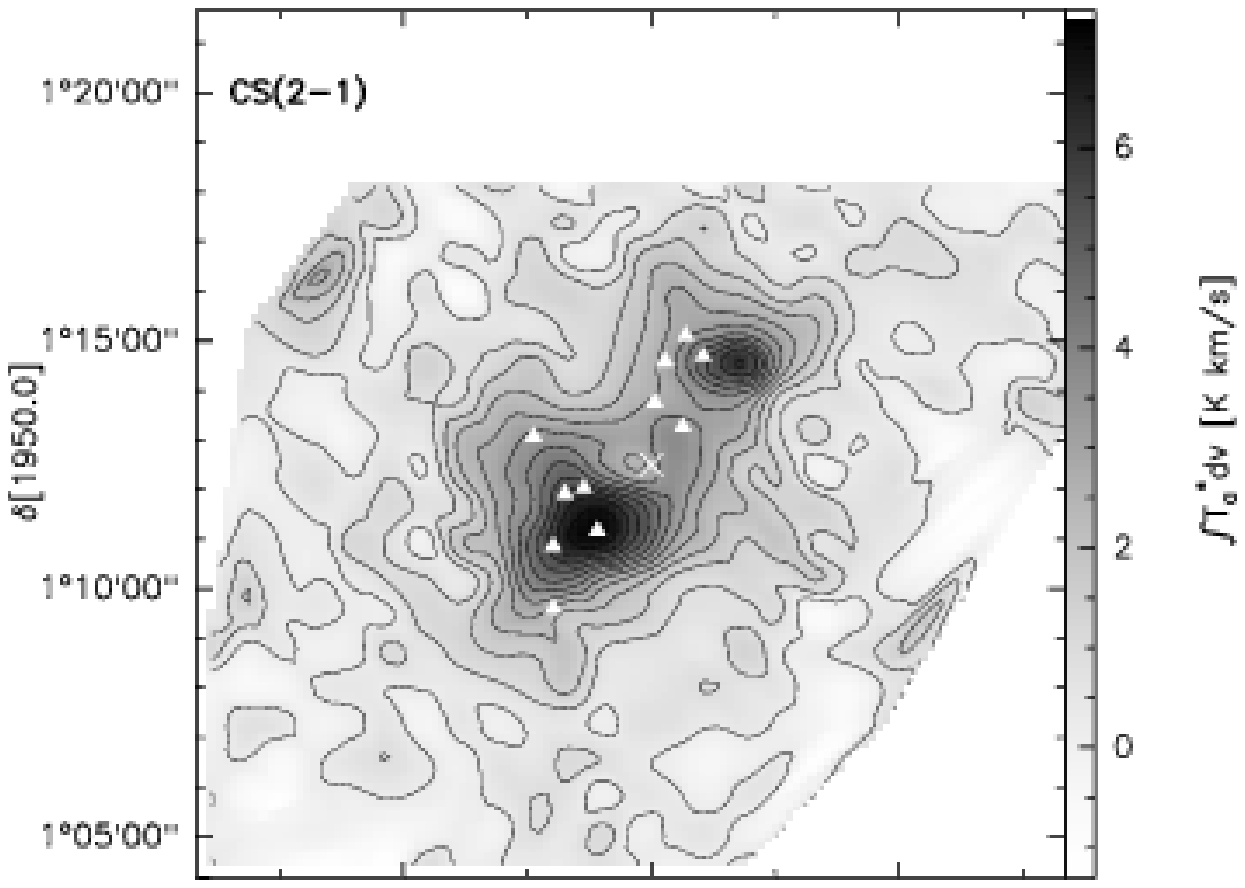,width=8.6cm,angle=270}}
\centerline{\psfig{figure=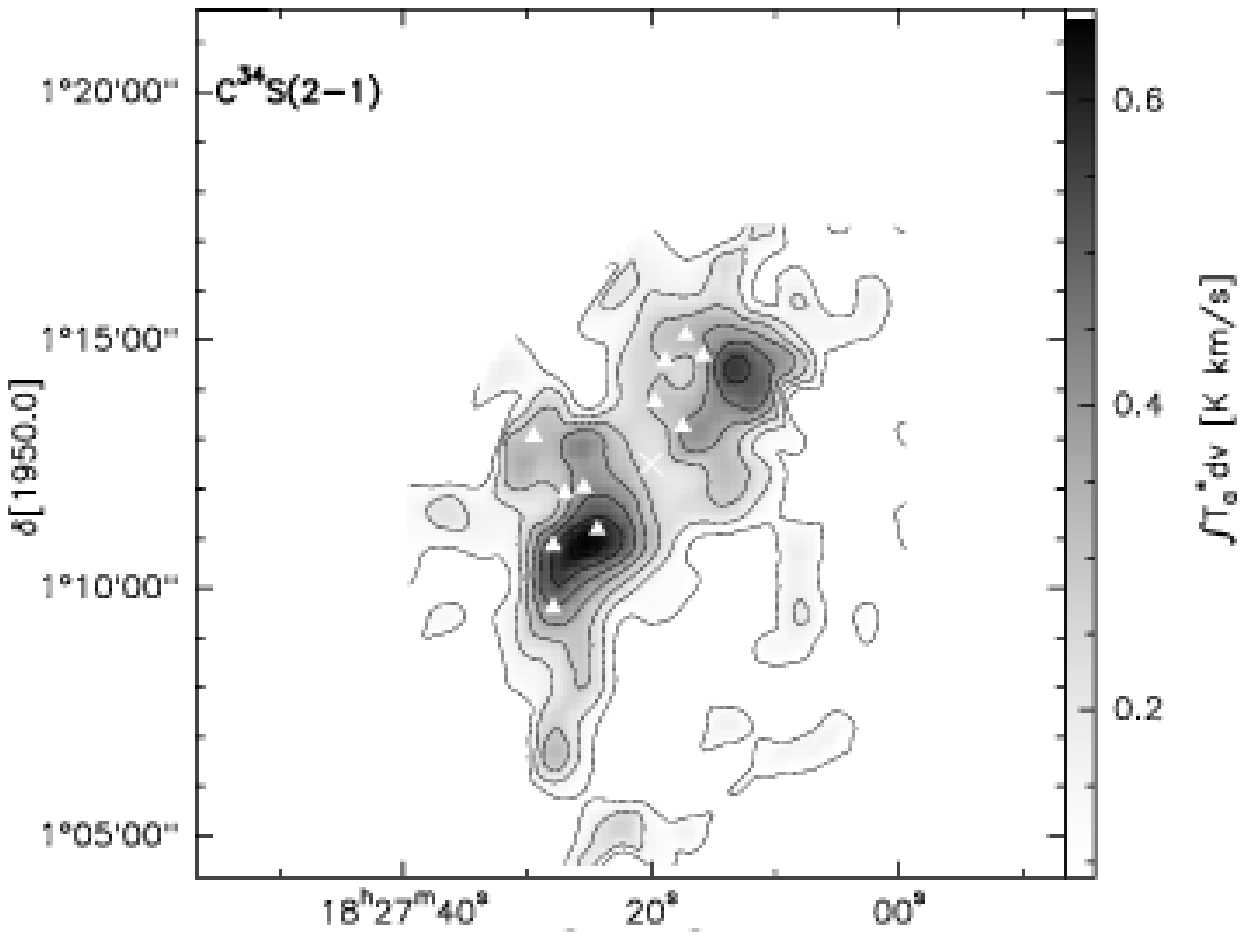,width=8.6cm,angle=270}}
\caption[ ]{
Maps of the CS and \CSI\ integrated intensity in the velocity
interval 2.0 to 13.0~\kms\ (CS) and 5.0 to 11.0~\kms\ (\CSI). 
In the CS map the value of the first
contour is 0.24~\Kkms and the contour interval is 0.6~\Kkms.
In the \CSI\ map the value of the first contour and the contour
interval are both 0.09~\Kkms.
}
\label{fig:csmap}
\end{figure}

\subsubsection{CS(2--1) and \CSI(2--1)}
\label{sec:csmorph}

To determine the distribution of the denser gas 
in Serpens we obtained integrated intensity maps 
of the CS and \CSI(2--1) emission (see also
Sect.\ref{sec:phpar}). The maps are presented in
Fig.~\ref{fig:csmap}, where CS and \CSI(2--1)
show several condensations and the two main clumps, NW and SE. 
A detailed comparison with the \N2H\ and \COII\ unveils several important
differences. TSOO
showed that the \N2H\ map follows the spatial distribution of the
submillimeter sources in both the NW and SE sub-clumps much more closely
than the other molecular tracers such as \CS\ and CH$_3$OH. This is confirmed
by our new single dish \CS\ and \CSI\ maps, showing that 
%
the CS distribution 
resembles much more closely that of other shock tracers, such as H$_2$CO 
(McMullin et al.~\cite{mcm00}), in agreement with what found by TSOO.
%
%

\subsection{Line profiles}
\label{sec:lineprof}
TSOO mapped the NW and SE clumps of the Serpens region in the \N2H(1--0)
transition and found four main cores of emission, A, B spatially associated  
with the SMM3/4 sources in the SE clump, and C, D associated with SMM1/S68N 
in the NW region. 
As mentioned earlier, they also
identified an extended, lower surface brightness Spur, stretching from the
NW to SE clumps. 
Spectra of \N2H\ and also of CS, \CSI\ and \COII\ at the emission peaks
of these regions are presented in Figs.~\ref{fig:abspt} to \ref{fig:spurspt}.
Our maps of the Serpens region obtained using CS, \CSI\ and \COII\
complement the \N2H\ observations of TSOO and show some new features.
In particular, the Spur is detected in all molecular tracers observed
by us and its emission only covers a narrow ($\simeq 0.4-0.5$~\kms)
velocity range. 
The \N2H(1--0) emission is mostly optically thin, except
at the emission peaks in the NW clump where $\tau({\rm N_2H^+})\la 1$. 

\subsubsection{\COI(1--0) and \COII(1--0)}
\label{sec:coline}

The \COI\ spectra in the NW clump appear to be self-absorbed (see
Fig.~\ref{fig:cdspt}), as the dip is at about the same velocity as
that of the CS spectra ($\simeq 8.6$~\kms) and is aligned with the
\CSI\ peak. The \COII\ spectra appear to have a Gaussian profile in
the NW clump and are mostly optically thin.  In this clump  the peak
of the \COII\ line is generally well aligned with those of \CSI\ and
\N2H\ and traces the dip of the \COI\ and CS spectra (see
Figs.~\ref{fig:cdspt} and \ref{fig:spurspt}).

\begin{figure}
%
%
%
\centerline{\psfig{figure=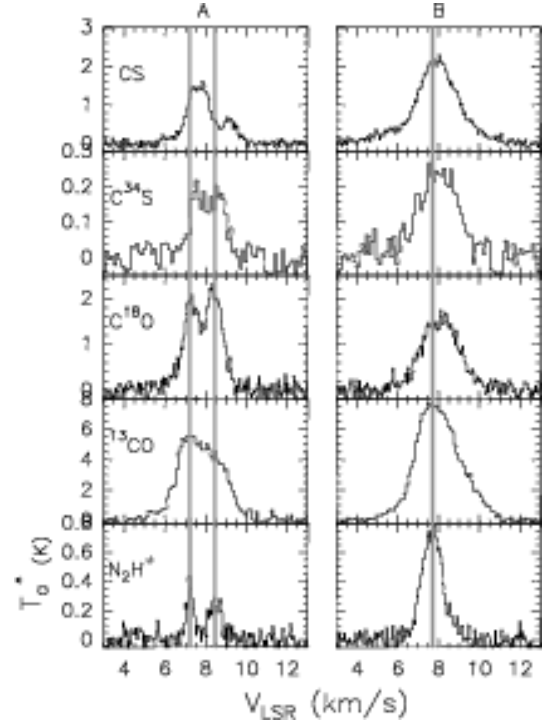,width=7.0cm,angle=0}}
\caption[ ]{
Line spectra at position A(133,-177) and B(88,-44) in the SE sub-cluster. From
top to bottom:
CS(2--1), \CSI(2--1), \COII(1--0), \COI(1--0), and \N2H(1$_{01}$--0$_{12}$).
The vertical lines are drawn at the peak velocities of the \N2H\ Gaussian fits. 
The highest velocity \N2H\ component corresponds to the
emission from the ``Spur''.
}
\label{fig:abspt}
\end{figure}

\begin{figure}
%
%
%
\centerline{\psfig{figure=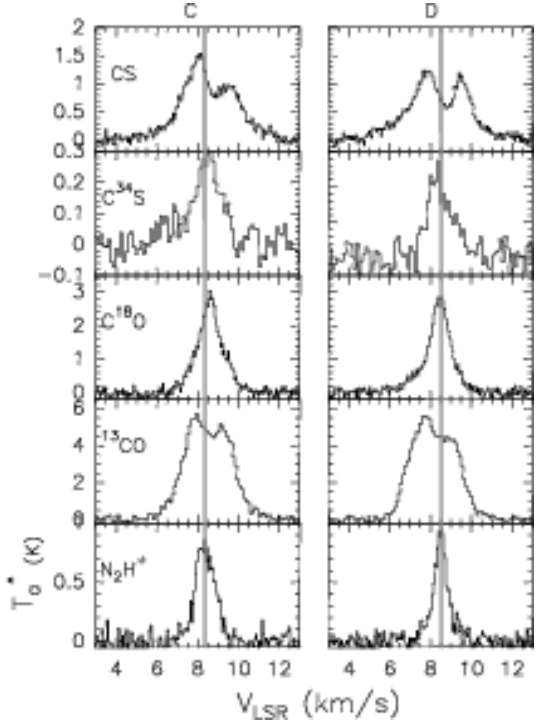,width=7.0cm,angle=0}}
\caption[ ]{
Same as Fig.~\ref{fig:abspt} for positions C(-44,44) 
(i.e., SMM1) and D(-44,133) (or S68N)
in the NW sub-cluster
}
\label{fig:cdspt}
\end{figure}

\begin{figure}
%
%
%
\centerline{\psfig{figure=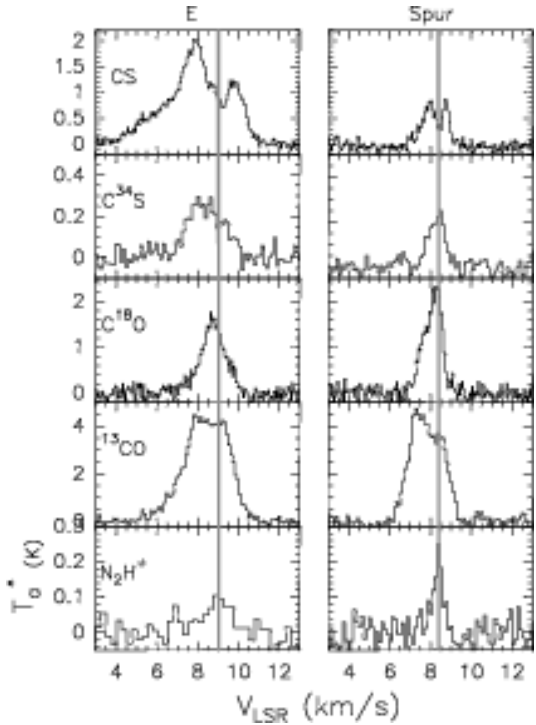,width=7.0cm,angle=0}}
\caption[ ]{
Same as Fig.~\ref{fig:abspt} for
positions E(-133,133) (or S68NW) and at one position in the Spur,
(133,-310)
}
\label{fig:spurspt}
\end{figure}

In the SE clump the line shape of \COI\ tends to be non-Gaussian,
especially at offsets $\delta<-100''$  and $\alpha>20''$ from the
map center. As for the \N2H\ line shape, the \COII\ non-Gaussian
profile is due to the  presence in the beam of several velocity
components (cores ``A'', ``B'' and the ``Spur'' of TSOO), as shown
in Fig.~\ref{fig:abspt} where the two \N2H\ peaks are nicely aligned
with those of \COII.  It is interesting to note that the ratio of
\COI\ to \COII\ emission 
is lower         
in the Spur 
than at the other positions.
For example at the position and velocity of core A, $(133,-177)$ and
$\vl=7.15$~\kms, the \COI\ to \COII\ intensity ratio is about 3, whereas at the
velocity of the Spur this ratio decreases to about 2.

\subsubsection{CS(2--1) and \CSI(2--1)}
\label{sec:csline}

Most of CS spectra in the NW clump
(see Fig.~\ref{fig:cdspt}) are self-absorbed and are optically thick, as
confirmed by the presence of the \CSI\ emission peaks at the dip of the CS
spectra (see also Williams and Myers~\cite{WM99}). The differences in the
profiles between the two isotopomers are mainly due to their different optical
depths: the \CSI\ lines are optically thin and single-peaked whereas the CS
spectra generally present two peaks. The asymmetries in the self-absorbed
CS spectra have been interpreted by Williams and Myers (\cite{WM99}) as the
signpost of a large-scale contraction of the gas around the clusters.
The blue asymmetric line profile of CS is visible in both the NW and SE
clumps, and by comparing CS and \CSI\ we will show later that in these
regions the centroid velocity of the optically thicker isotopomer is
lower (bluer) than its rarer and optically thin counterpart.

The CS spectra show line wings due to outflows from  several cluster
sources, such as S68N and SMM1 in the NW clump, and SMM3 and SMM4 in
the SE clump. The CS spectra in the Spur are self-absorbed, as
shown in  Fig.~\ref{fig:spurspt}, and both the \N2H\ and \COII\ emission
peaks are aligned with the dip of the CS spectra.

\subsection{Kinematics of the Serpens region}
\label{sec:kin}

The analysis of the gas kinematics in the Serpens region is complicated
by the presence of several velocity fields that may include infall, 
rotation, turbulence and outflow motions. In particular, the presence of 
several known outflows may render the unambiguous detection of either
rotation or infall very difficult. In this work we did not attempt
to detect infall  motions around some of the specific submillimeter
sources, but we were instead interested in the {\it global} motions 
of the cloud and within the cloud. The molecular tracers used also 
reflect this goal: their excitation conditions make \COI\ and 
\COII(1--0) less prone to contamination by outflows. Likewise, the two 
optically thin tracers \CSI(2--1) and \N2H(1--0) are not affected
by the presence of outflows. However, the CS(2--1) emission is known
to be contaminated in the presence of strong outflows (see TSOO and
refrences therein) and indeed we detect line wings in the CS lines 
(see Sect.~\ref{sec:csline}).

We present the velocity information in four different 
graphical forms designed to emphasize different kinematical properties 
of the gas:
(i) line core, red and blue wing maps to show large-scale velocity structures;
(ii) centroid velocity maps, to show small velocity trends of the cloud
bulk emission; (iii) normalized centroid velocity difference between an optically
thick and a thin tracer, to show systematic differences such as infall motions;
(iv) line width maps, to show the velocity dispersion accross the region for
the various tracers.

\subsubsection{Line core and line wings maps}
\label{sec:linec}

\begin{figure*}
%
%
\vspace*{6mm}
\centerline{\psfig{figure=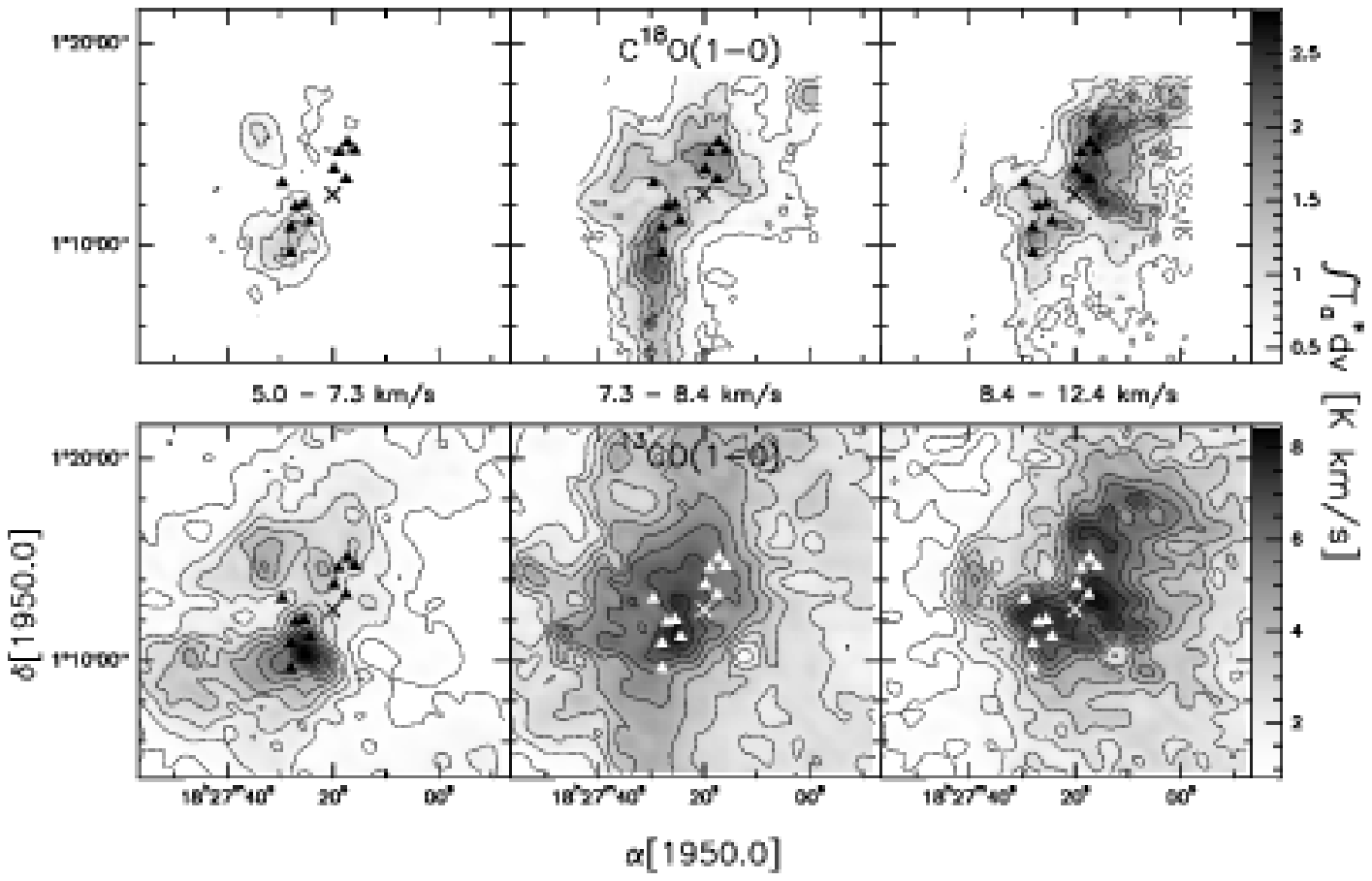,width=15.0cm,angle=270}}
\centerline{\psfig{figure=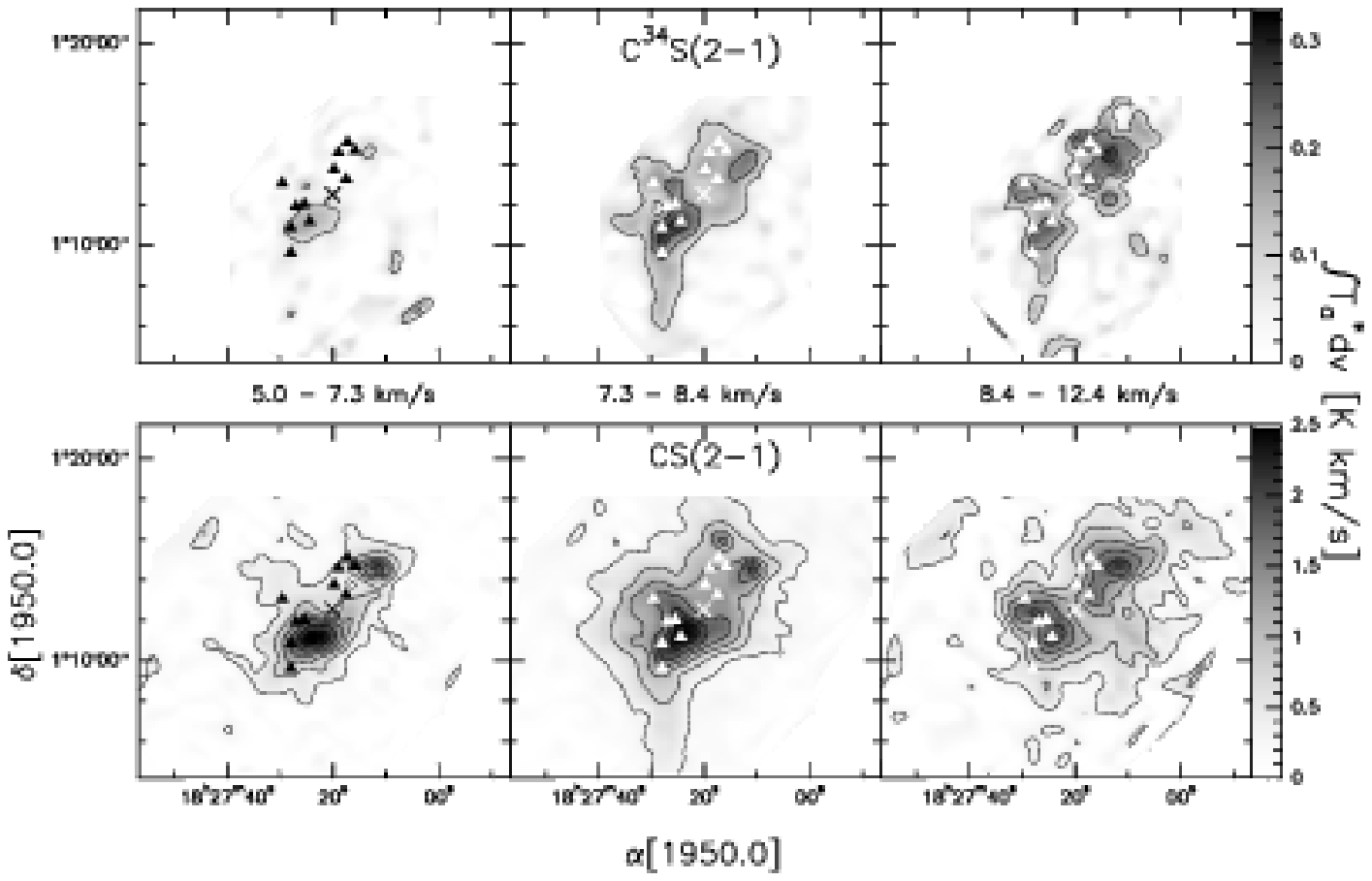,width=15.0cm,angle=270}}
\caption[ ]{From top to bottom:
line core and line wings integrated intensity maps for
\COII, \COI, \CSI, and \CS.
}
\label{fig:coch}
\end{figure*}

To determine the velocity distribution of the ambient gas in Serpens we
made integrated intensity maps of the blueshifted ($5.0-7.3$~\kms), 
line core ($7.3-8.4$~\kms) and redshifted ($8.4-12.4$~\kms)
emission of \COI\ and \COII(1--0). The maps are shown in
Fig.~\ref{fig:coch}  where the \COI\ and \COII\  emission is clearly
separated and occupies mainly the half-plane to the East in the blueshifted
channel, whereas the emission is mostly concentrated in the half-plane
to the West in the redshifted channel (see also 
Fig.~\ref{fig:13cocv}). In particular, \COI\ shows a very different 
distribution from that of CO(1--0) (McMullin et al.~\cite{mcm94}) 
and (2--1) (Davis et al.~\cite{dav99}). In fact, while the CO emission 
is affected by the high-velocity gas radiated from the submillimeter sources,
\COI\ is less affected by molecular ouflows and thus can better show the 
kinematics of the large-scale gas.
In the central velocity map of \COII\ the Spur is clearly visible as a 
gas filament extending southward from the SE cluster of submillimeter sources.

The maps in Fig.~\ref{fig:coch} 
show emission from the SE and NW sub-clusters at all velocities, although
the \CSI\ emission is weak in the $5.0-7.3$~\kms velocity interval and
is mostly concentrated in the SE clump. We also note that the \COII(1--0)
emission from the NW clump is concentrated in the line core and especially 
in the redshifted channel. The morphology shown by the channel maps 
of the two CO isotopomers and of the density sensitive tracers is clearly
different. The \COI\ and \COII(1--0) line emission show a velocity gradient
roughly E to W whereas CS(2--1) is tracing the embedded denser cloud cores
and the EW velocity gradient is much less evident.

Moreover, they show other isolated cores of emission, such as S68N, visible
in the CS central velocity map. This core has a weak \N2H\ counterpart
shown in the \N2H\ channel map of TSOO at a velocity of about 8.2~\kms.
Another emission peak at the position of SMM1 is visible in both the CS
and \CSI\ 8.4--12.4~\kms channel maps; in \CSI\ this emission peak is also
visible in the other velocity channels.  The Spur is also visible in the
central velocity map. Finally, in the CS 8.4--12.4~\kms
channel map the SE sub-cluster appears to be composed of two separate cores,
one at the position of SMM4 and the other slightly north of the SMM 3/6
group.

%
%

The central velocity \COI\ map shows an emission peak at the position of SMM4
and of the CS peak. Another peak is visible sligthly NW of SMM3; the \N2H\
core B is positioned between these two emission peaks. In the 5.0-7.3~\kms
channel map a sharp peak of \COI\ emission is visible South of SMM4, which is
also the region of \COII\ emission, although the peaks are not coincident.
A secondary peak of \COII\ emission can be found in the NE region of this
velocity channel that is coincident with enhanced \COI\ emission. There are
clearly two velocity components overlapping at this position, as shown by
the \COII\ and \COI\ spectra.

\subsubsection{Large-scale velocity gradient}       
\label{sec:velfit}


The velocity gradient of an interstellar gas cloud can be
determined by using all or most of the data in a map at once,
by least-squares fitting maps of line-center velocity for 
the direction and magnitude of the best-fit velocity gradient
(see Goodman et al. \cite{good93} and references therein).

A cloud undergoing solid-body rotation would exhibit a linear
gradient, ${\rm d}V/{\rm d}r$, across the face of a map, perpendicular
to the rotation axis. Thus, following Goodman et al. (\cite{good93})
we fit the function
\beq
\Vlsr = V_o+a\Delta\alpha+b\Delta\delta
\label{eq:vfit}
\eeq
to the peak velocity of a Gaussian fit to the emission profile 
of several molecular tracers. Here, $\Delta\alpha$ and $\Delta\delta$
represent offsets in right ascension and declination, expressed in
radians, $a$ and $b$ are the projections of the gradient per radian 
on the $\alpha$ and $\delta$ axes, respectively, and $V_o$ is the LSR 
systemic velocity of the cloud.
The magnitude of the velocity gradient, in a cloud at distance $d$,
is thus given by
\beq
\frac{{\rm d}V}{{\rm d}r} = \frac{(a^2+b^2)^{1/2}}{d}
\eeq
and its direction (measured east of north) is
\beq
\theta_{\rm v} = \tan^{-1}\frac{a}{b}
\eeq
In order to carry out a least-squares fit to Eq.(\ref{eq:vfit}) 
we have computed the following error function
\beq
\chi^2 = \sum_{i,j}{} \left [ 
\frac{\Vlsr(i,j)-V_o-a\Delta\alpha_i-b\Delta\delta_j}{\sigma_v(i,j)}
\right ]^2
\label{eq:chi}
\eeq
where $\Vlsr(i,j)$ is the velocity of the Gaussian fit to the spectrum
at position $(\Delta\alpha_i,\Delta\delta_j)$ in the map, and 
$\sigma_v(i,j)$ is the uncertainty in $\Vlsr$ determined by the fit,
given by (see Goodman et al. \cite{good93} and references therein):
\beq
\sigma_v = 1.15 \left ( \frac{\sigma_T}{T} \right ) 
(\delta_v \Delta V)^{1/2}
\eeq
where $T$ is the peak of a Gaussian fit to the line profile, $\sigma_T$
is the RMS noise in the spectrum, $ \delta_v$ is the velocity resolution of 
the spectrum and $\Delta V$ is the FWHM line width. Only those points
for which SNR$=T/\sigma_T>3$ were used in Eq.(\ref{eq:chi}).

To find the minimum of the function expressed by Eq.(\ref{eq:chi}) we have 
used the downhill simplex method, using ``AMOEBA'' of Press et al. 
(\cite{nr92}) as the basic multidimensional minimization routine. The
results are listed in Table~\ref{tab:vfit} for the various molecular
tracers. Two comments are in order: {(i)} \N2H\ and \CSI\ trace
the denser gas in the cores and are thus less suited to calculate the 
{\it large-scale} velocity gradient; {(ii)} \COI\ is mostly optically thick
with a non-Gaussian line profile and it is also not reliable. 
For these reasons we choose \COII\ as the best available
tracer to us of the velocity gradient in the Serpens cloud. 
Despite the aforementioned {\it caveats}, the direction of the velocity 
gradient is not significantly different from one tracer to another 
and confirms our early assumption in Sect.~\ref{sec:linec} that the 
Serpens cloud exhibits a velocity gradient roughly E to W.

%
%
%
\begin{table}
\begin{flushleft}
\caption[ ]{
Results of velocity gradient fitting. 
}
\begin{tabular}{lccr}
\hline
Line   & $V_o$   & ${\rm d}V/{\rm d}r$     & $\theta_v$ \\
       & [\kms]  & [\kms~pc$^{-1}$]     & [deg E of N]  \\
\hline
\COII(1--0)                & 8.31   & 1.25  & -88.9 \\
\COI(1--0)                 & 8.03   & 0.75  & -70.8 \\
\CSI(2--1)                 & 8.42   & 2.03  & -78.9 \\
\N2H(1$_{23}$--0$_{12}$)   & 8.34   & 0.54  & -72.4 \\
\hline
\end{tabular}
\label{tab:vfit}
\end{flushleft}

\end{table}


\subsubsection{Centroid velocity maps}

The centroid velocity of a line is that velocity at which the
integrated intensity is equal on either side of the line profile, and we
compute it as in Narayanan et al. (\cite{nar98}). Centroid velocity
maps are usually a good indicator of global velocity structures, such
as rotational motions, and 
we use them here as further evidence of
the large-scale velocity gradient found in the previous section. 

\begin{figure}
%
%
%
%
%
\vspace*{4mm}
\centerline{\psfig{figure=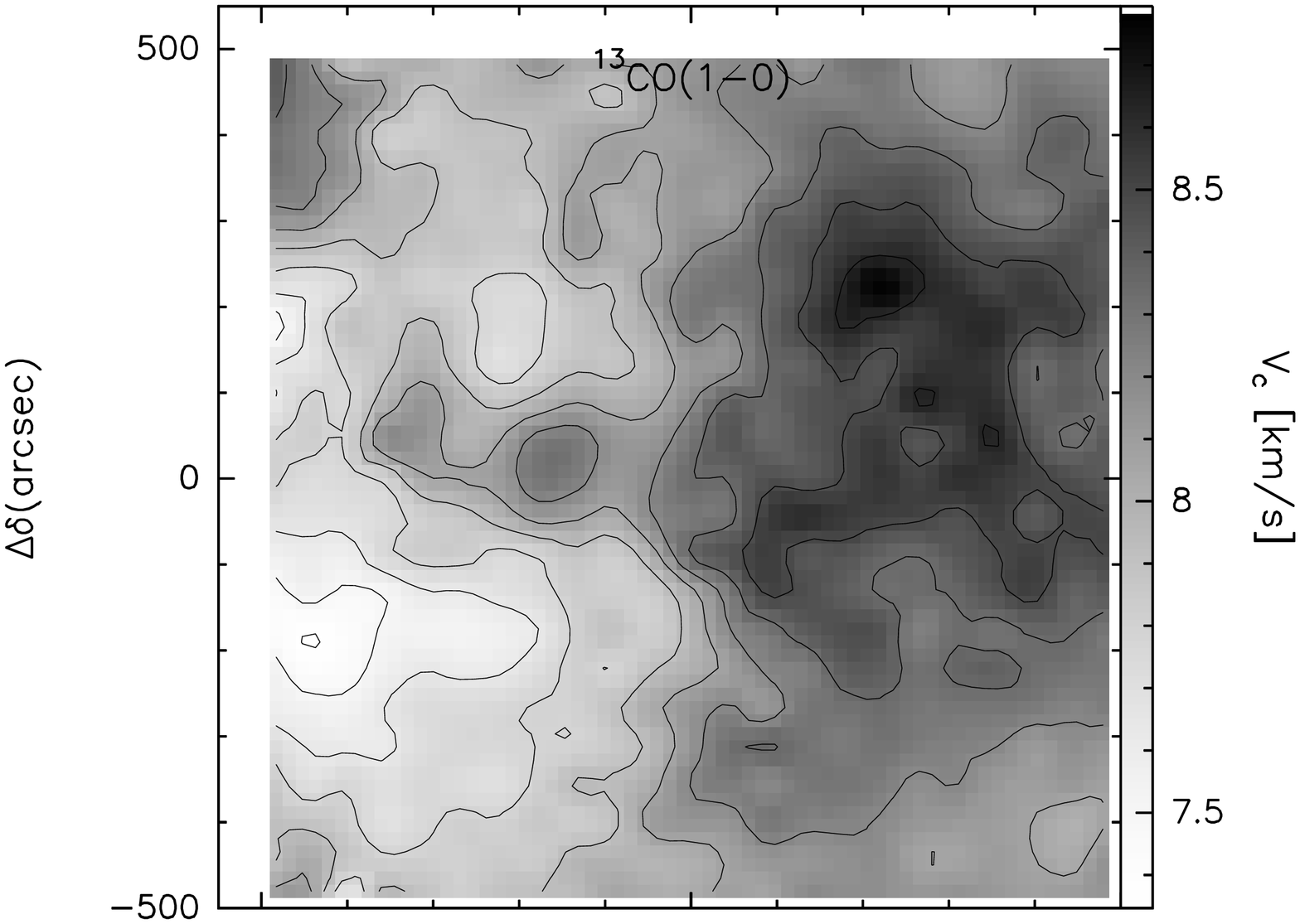,width=7.0cm,angle=270}}
\centerline{\psfig{figure=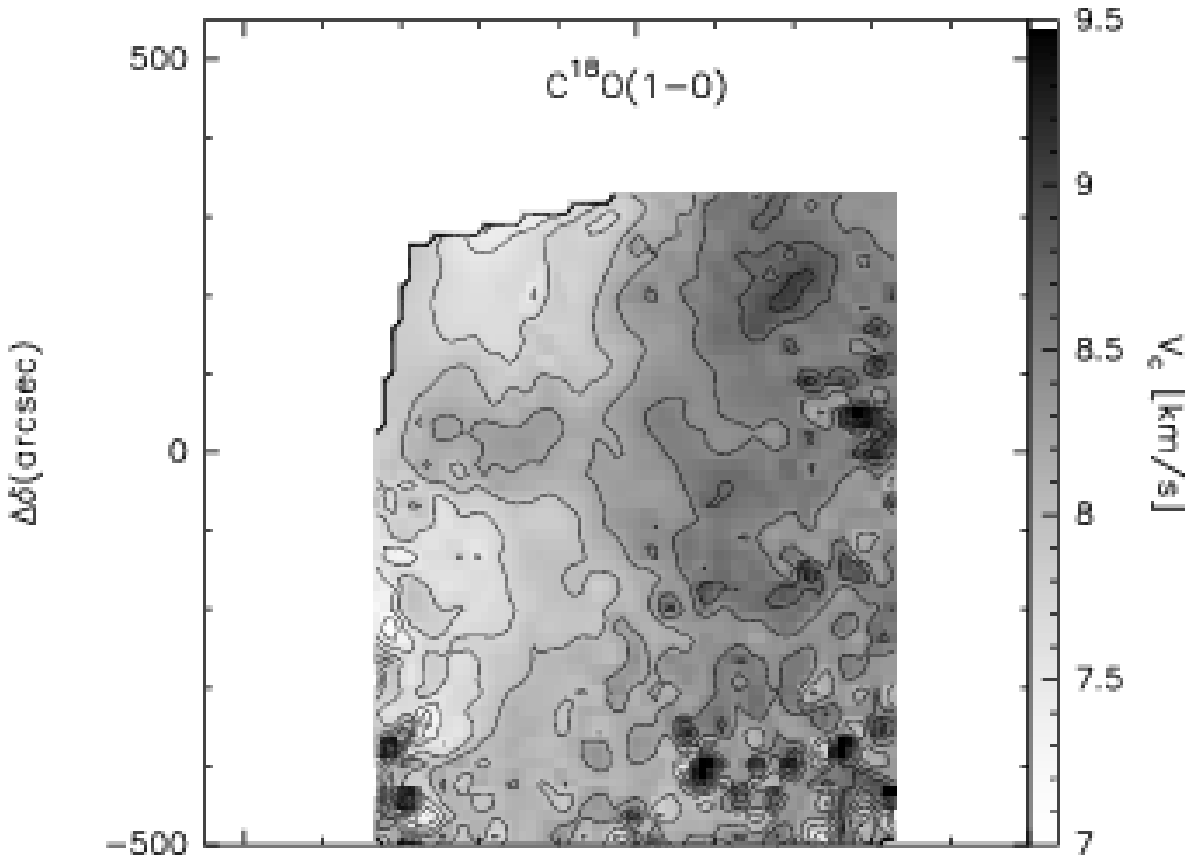,width=6.8cm,angle=270}}
\centerline{\psfig{figure=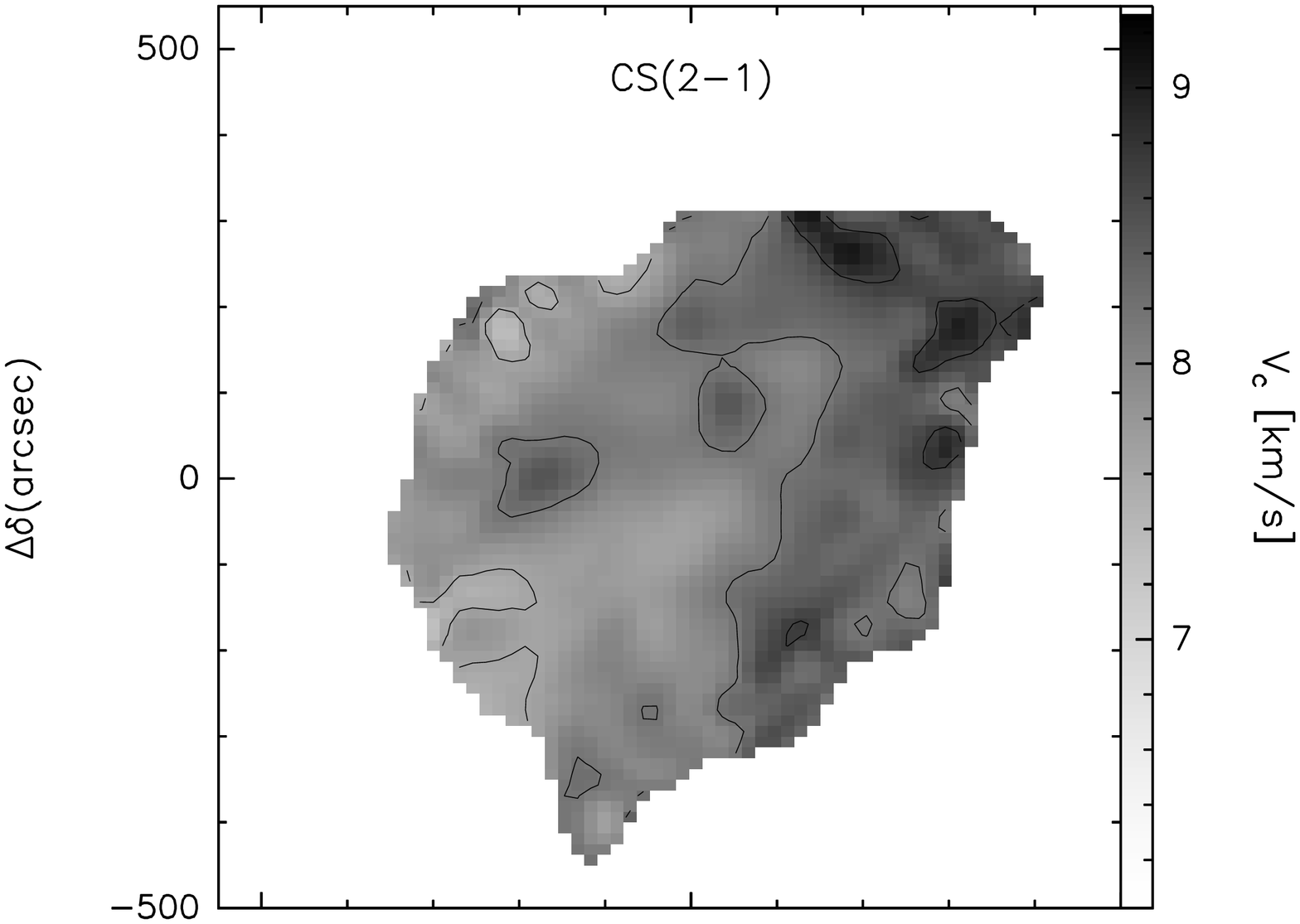,width=7.0cm,angle=270}}
\vspace*{-0.45cm}
\centerline{\psfig{figure=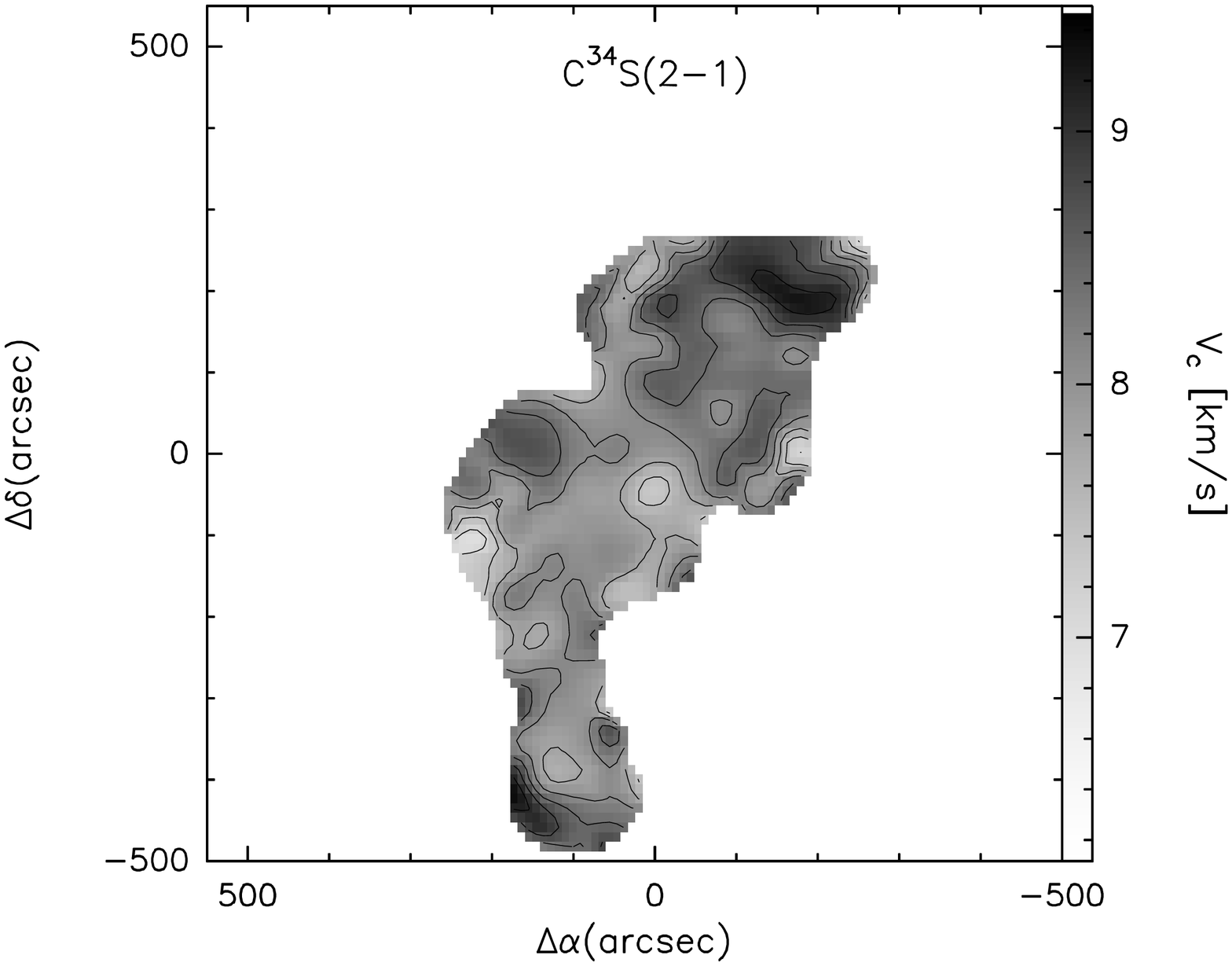,width=7.0cm,angle=270}}
\vspace*{0.5cm}
\caption[ ]{
Centroid velocity maps for: \COI, \COII, \CS, and \CSI\ (from top to bottom).
The centroid velocities are
obtained over the velocity interval corresponding to the line core
(6 to 10.5 \kms). Only the observed points where the integrated intensity
is greater than three times its corresponding RMS uncertainty are
used in calculating the centroid velocity. 
The offsets are measured with respect to the center position listed 
in Table~\ref{tab:lines}. 
%
}
\label{fig:13cocv}
\end{figure}

In Fig.~\ref{fig:13cocv} we present the centroid
velocity maps of the \COI\ and
\COII\ lines. To minimize the potential contamination of the kinematics of
the cloud by the outflow velocity fields, the centroid velocities were
computed over line core intervals ($6.0-10.5$~\kms) which must also be wide
enough to encompass the line core emission in both the NW and SE sub-clumps.
Both maps show an increasing gradient in centroid velocity from roughly E to
W, as expected. 

The maps shown in Fig.~\ref{fig:13cocv} 
also show a velocity ``bulge'' extending
from about the map center towards the E, which may be caused by one of
the redshifted CO(1--0) outflow lobes found by Narayanan et al. 
(\cite{nmwb01}), despite our centroid velocities are calculated
using only line core emission. Even if some contamination of
the centroid velocity maps by the outflows may still be present, the
velocity gradient observed in the maps of Fig.~\ref{fig:13cocv} 
is extended
to the whole Serpens cloud and is unlikely that the \COI\ {\it and} 
\COII\ centroid velocities are seriously affected by the outflows.
We can also exclude that the observed velocity gradient is due to 
separate clumps with different systemic velocities, as the differences
in the lines $\vl$ are less or comparable with the line FWHM of \COII\
and show a smooth change from E to W. As will be further discussed in
Sect.~\ref{sec:rot}, we conclude
that the observed velocity gradient may indeed be caused
by a global rotation of the Serpens molecular cloud whose rotation axis
is roughly aligned in the SN direction.

In Fig.~\ref{fig:13cocv} we also
 present the centroid velocity maps of the CS and
\CSI(2--1) transitions. The CS map shows essentially a plateau in the 
region where most of the CS(2--1) emission is detected. Our spatial 
resolution is not good enough to distinguish any particular feature around
specific submillimeter sources.  However, we note once again the 
small bulge of redshifted emission E of the map center. 


\subsubsection{Normalized velocity difference maps and infall speeds}
\label{sec:normvel}

A diagnostics which is often used to disentangle the large-scale 
kinematics of molecular cloud cores is based on the
normalized centroid velocity difference (see Mardones et al. 
\cite{mard97}).
The normalized centroid velocity difference $\delta V_{\rm c}$ is
a non-dimensional parameter defined as:
\beq
\delta V_{\rm c} = \frac{V_{\rm c}({\rm CS}) - V_{\rm c}({\rm C^{34}S})}   
{\Delta V({\rm C^{34}S})}
\label{eq:diffcv}
\eeq
and is very useful to quantify the observed line asymmetries in the optically thick
CS(2--1) line.  Alternatively, one can use \N2H(1--0)
to replace \CSI\ as the optically thin tracer in Eq.~(\ref{eq:diffcv}).
A blue asymmetry ($\delta V_{\rm c}\le 0$) can be interpreted as an indication
of infall motion (Myers et al.~\cite{Mea96}; Mardones et al.~\cite{mard97}).

\begin{figure}
%
%
\centerline{\psfig{figure=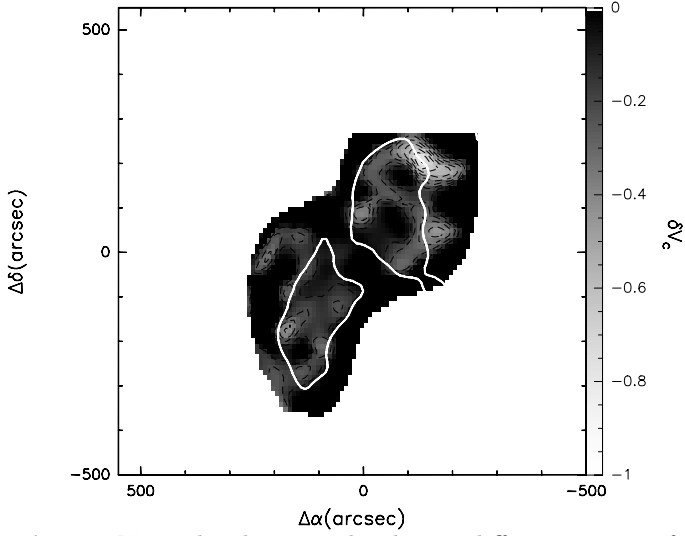,width=9.0cm,angle=270}}
\caption[ ]{
Normalized centroid velocity difference map of CS and \CSI. 
The centroid velocities were
calculated over the velocity interval 5.5 to 10.5 \kms. To better
show the contrast between the regions where $\delta V_{\rm c}<0$ and 
those where $\delta V_{\rm c}>0$ the upper limit for this map 
has been set equal to 0. The thick white line represents the contour
for the 50\% of the peak of the \HM\ column density.
}
\label{fig:diffcv}
\end{figure}

In Fig.~\ref{fig:diffcv} the map 
of $\delta V_{\rm c}$ as defined in Eq.~(\ref{eq:diffcv}) is shown. 
We decided to use CS as the optically thick tracer and \CSI, rather 
than \N2H, as the optically thin tracer to ensure that both species 
have homogeneous chemical properties. We can see that there are two 
regions of the map where the negative values of $\delta V_{\rm c}$ 
are more concentrated, which are approximately coincident with the 
SE and NW sub-clumps. This indicates that there is molecular gas
around the SE and NW sub-clusters which is still undergoing an 
infalling motion. 

Typical infall speeds at different positions in the SE and NW clumps
can be calculated by using the simple two layer model of Myers 
et al. (\cite{Mea96}), which gives:
%

\beq
V_{\rm in} \simeq \frac{\sigma_{\rm v}}{V_{\rm red}-V_{\rm blue}} 
\log \left ( \frac{1+eT_{\rm BD}/T_{\rm D}}{1+eT_{\rm RD}/T_{\rm D}}
\right )
\eeq
where $T_{\rm D}$ is the brightness temperature of the dip in the CS line
shape, $T_{\rm BD}$ is the height of the blue peak above the dip, and
$T_{\rm RD}$ is the height of the red peak above the dip. The velocity
dispersion $\sigma_{\rm v}$ can be obtained from the FWHM of an optically 
thin line, such as \N2H, thus 
$\sigma_{\rm v}=\Delta V_{\rm N_2H^+}/\sqrt{8 \log 2}$. The results are
shown in Table~\ref{tab:vin}, where the first two positions refer to
the SE clump, and the next three correspond to the NW clump. 
In the SE clump (including position B, not shown in Table~\ref{tab:vin})
the CS line profiles are more symmetric, implying a lower infall speed,
except near position A where $V_{\rm in}>0.2$~\kms. This suggests  
different states in the core contraction process, as also shown by
Fig.~\ref{fig:diffcv}. 

%
%
%
\begin{table}
\begin{flushleft}
\caption[ ]{
{\bf Line velocity dispersion of \N2H\ (second column), non-thermal
velocity dispersion (third column) and infall speeds (fourth column) 
towards the SE and NW sub-clumps ($\sigma=\Delta V/\sqrt{8 \log 2}$).}
}
\begin{tabular}{lccr}
\hline
Position   & $\sigma_{\rm v}$  & $\sigma_{\rm NT}$   & $V_{\rm in}$   \\
           & [\kms]            & [\kms]              & [\kms]       \\
\hline
A          & 0.33      & 0.32         & 0.26  \\
(88,-177)  & 0.35      & 0.34         & 0.21  \\
\hline
C          & 0.49      & 0.48         & 0.35  \\
D          & 0.35      & 0.34         & 0.09  \\
E          & 0.62      & 0.58         & 0.44  \\
\hline
\end{tabular}
\label{tab:vin}
\end{flushleft}

\end{table}

A similar difference clearly exists in the NW sub-clump, where positions
C and E show substantially different infall speed compared to 
position D. This difference has also been observed by Williams \& Myers   
(\cite{WM99}), and our results confirm their conclusion 
about the presence of supersonic
($V_{\rm in}>\sigma_{\rm th}[{\rm H_2}]=0.27$~\kms at $T=20$~K)
inward motions in the NW sub-clump, and indicate near-supersonic 
contraction in the SE sub-clump.

For these same positions we can quantify the turbulent motion by comparing
the \N2H\ line FWHM with that expected for gas having equal thermal and 
non-thermal motions (Mardones et al.~\cite{mard97}, Lee et al.~\cite{L99}):
\beq
\Delta V_{\circ} = \left [ 8 \log 2 \, k T \left ( \frac{1}{m_{\rm N_2H^+}}
+ \frac{1}{m_{\rm av}} \right ) \right ]^{1/2}
\eeq
where T is the gas kinetic temperature, $k$ is the Boltzmann constant, 
$m_{\rm N_2H^+}$ is the molecular
weight of \N2H\ (29 amu), and $m_{\rm av}$ is the gas mean molecular 
weight (2.3 amu). For gas with $T=20$~K, we find that 
$\Delta V_{\rm N_2H^+} > \Delta V_{\circ} = 0.66$~\kms, thus having
greater non-thermal than thermal motions (see Table~\ref{tab:vin}). 
This result is consistent 
with the prevalence of turbulent infall motions in a sample of cores
with embedded YSOs found by Mardones et al.
(\cite{mard97}), whereas  Lee et al. (\cite{L99}) found that most 
starless infall candidates are in a state of thermal infall.
Incidentally, we note that also the starless core at position E 
has a greater turbulent rather than thermal motion.

The aim of this paper is not to calculate the mass infall rate 
for every cloud core, as our spatial resolution is not high enough.
However, for consistency we did calculate the kinematic mass infall
rate, 
$({\rm d}M/{\rm d}t)_{\rm kin}=
4\pi R_{\rm in}^2 \bra m_{\rm av} n_{\rm H_2} \ket V_{\rm in}$,
toward position E. The infall radius was taken from Williams \& Myers 
(\cite{WM99}), $R_{\rm in}=6200$~AU, $V_{\rm in}=0.44$~\kms\ from 
Table~\ref{tab:vin}, and $\bra m_{\rm av} n_{\rm H_2} \ket $ 
was obtained by taking
the average of all mass estimates in Table~\ref{tab:masses} (except for the 
values obtained from the statistical equilibrium method, 
see Sect.~\ref{sec:dens}) and dividing by the total volume of the NW clump. 
Thus, we obtain 
$({\rm d}M/{\rm d}t)_{\rm kin}\simeq 5 \, 10^{-5}$~\solmass yr$^{-1}$. 
This value can be compared with the gravitational mass accretion rate 
of $({\rm d}M/{\rm d}t)_{\rm gr}=0.975 a^3/G$ (Shu \cite{shu77}), 
where $a$ is the velocity dispersion
of the molecule of mean mass $m_{\rm av}$, 
$a^2 = kT/m_{\rm av} + \sigma_{\rm NT}^2$. Using $T=20$~K and $\sigma_{\rm NT}$
from Table~\ref{tab:vin}, we find 
$({\rm d}M/{\rm d}t)_{\rm gr}\simeq 5.9\, 10^{-5}$~\solmass yr$^{-1}$.
The agreement of these two rates indicates that the derived inward 
motion at position E is consistent with gravitational infall.


\subsubsection{Line width maps}

\begin{figure}
%
%
%
%
\centerline{\psfig{figure=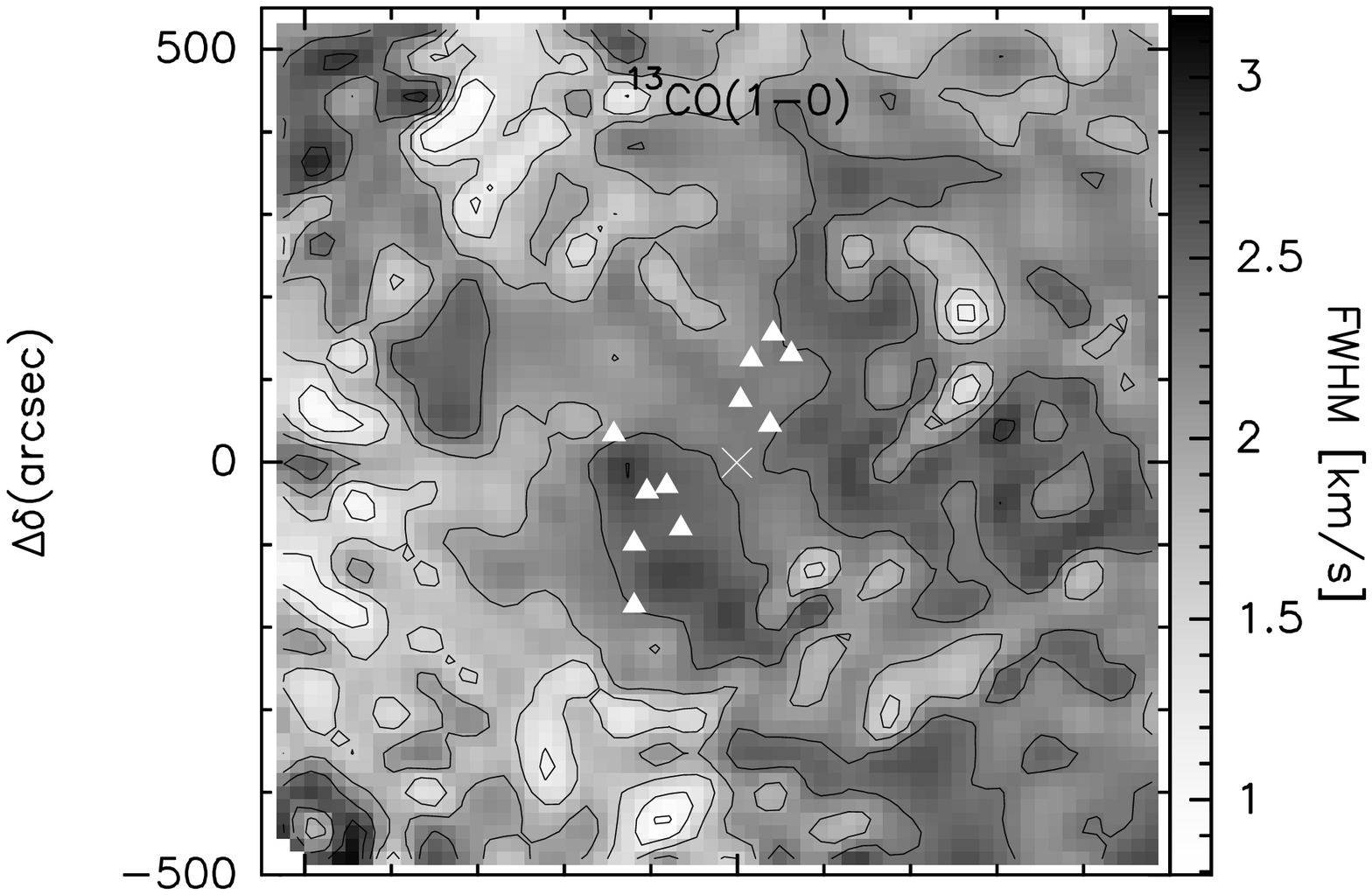,width=7.0cm,angle=270}}
\centerline{\psfig{figure=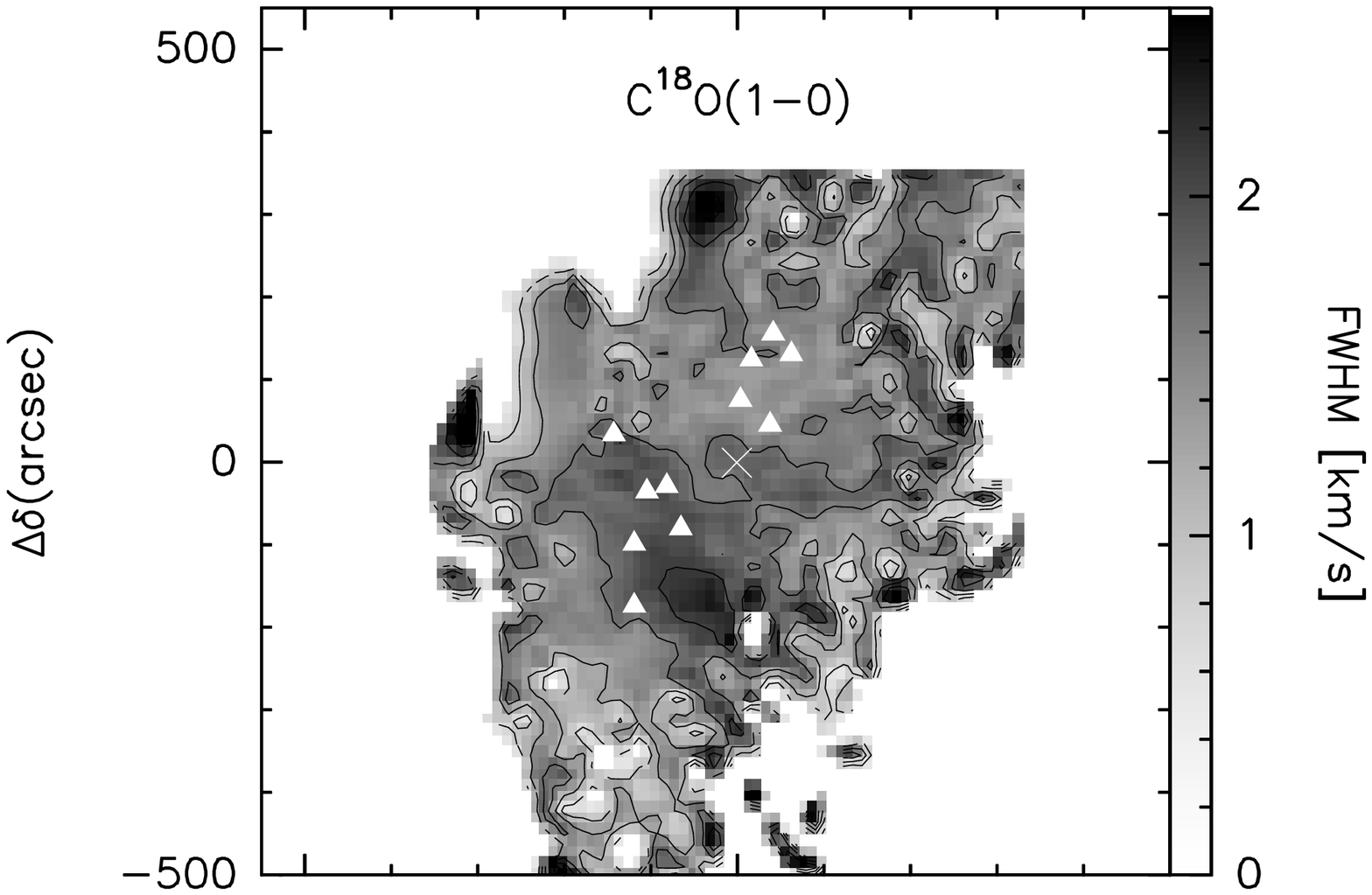,width=7.0cm,angle=270}}
\centerline{\psfig{figure=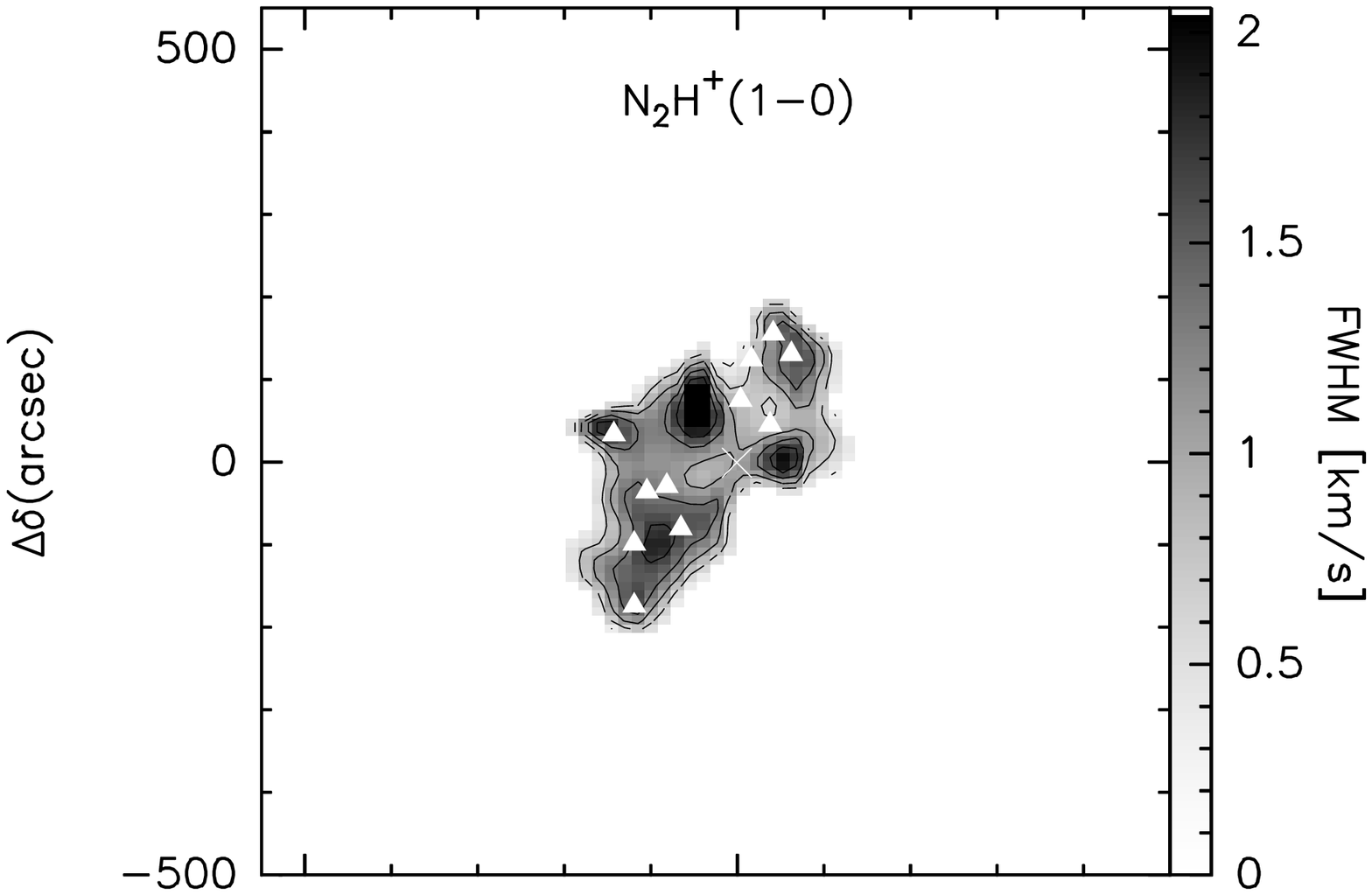,width=7.0cm,angle=270}}
\centerline{\psfig{figure=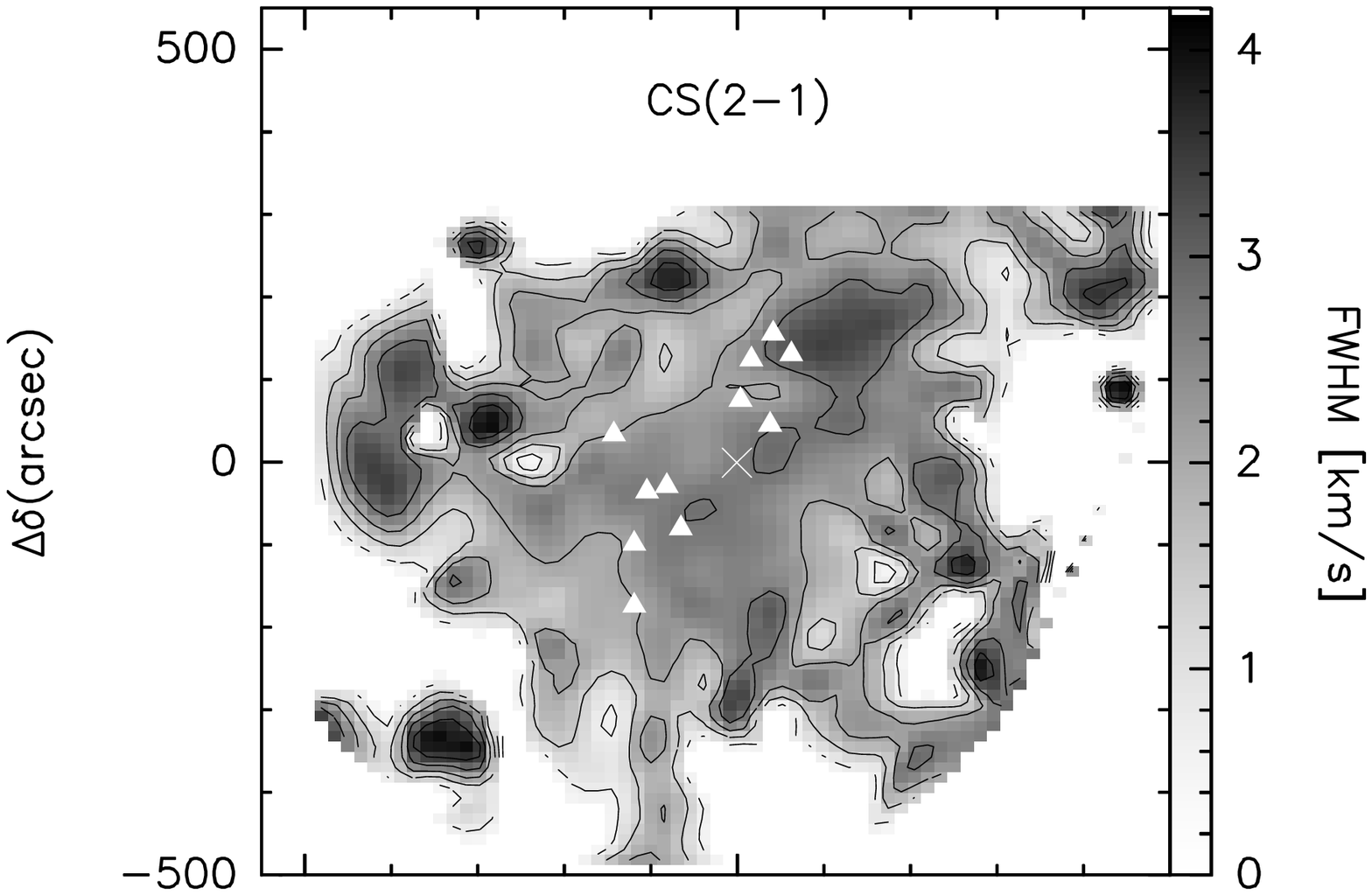,width=7.0cm,angle=270}}
\centerline{\psfig{figure=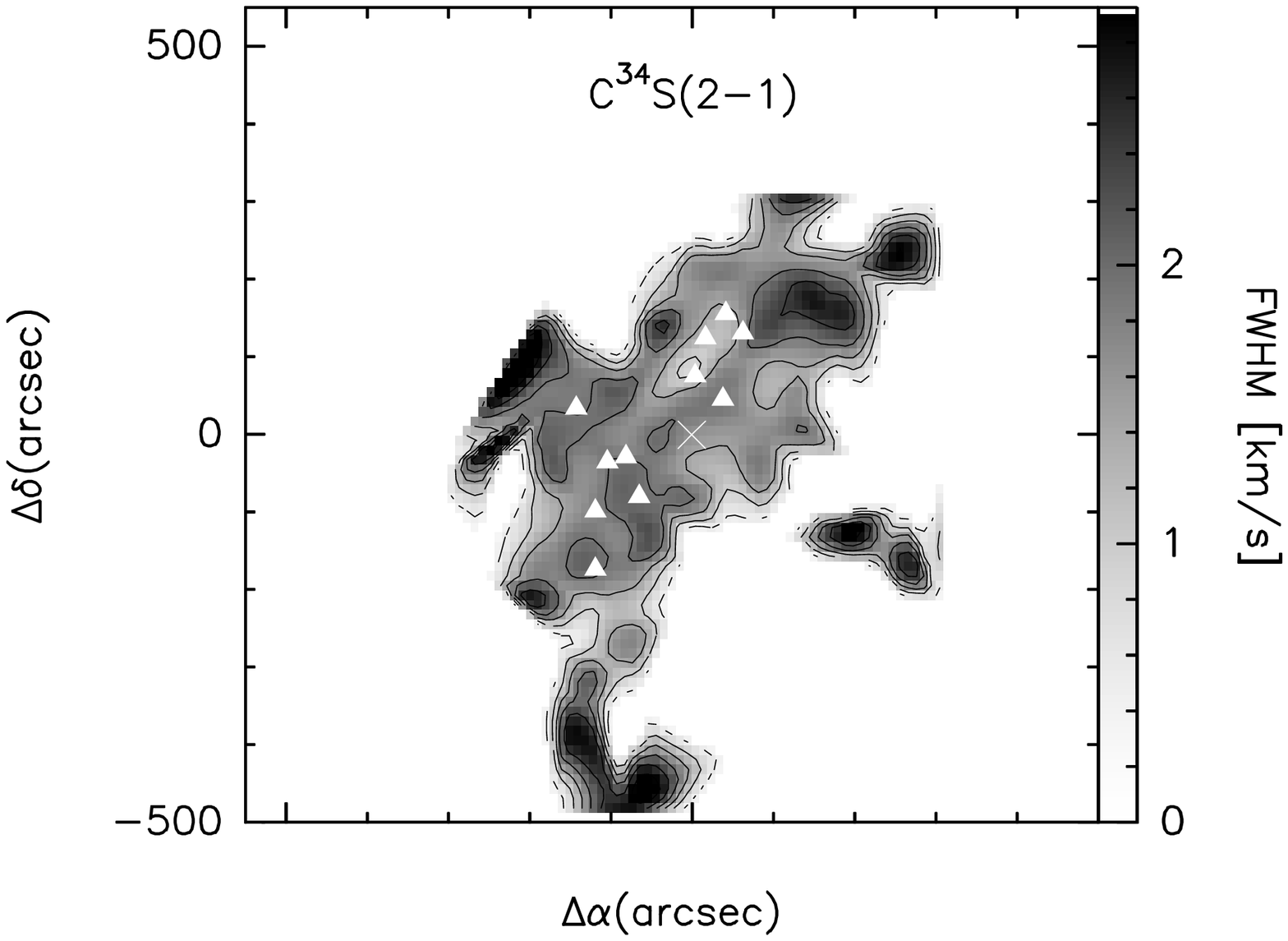,width=7.0cm,angle=270}}
\vspace*{5mm}
\caption[ ]{
Maps of the $2^{nd}$ moment of the data (line width). From
top to bottom: \COI, \COII, \N2H, \CS, \CSI. Data with SNR$<5\sigma$
have been masked out. The \N2H\ map shows the  $2^{nd}$ moment of
the first hyperfine component.
}
\label{fig:cowidth}
\end{figure}

Another useful diagnostic tool is to plot a measure of the width
of the line profile. Given the deviation from a Gaussian line profile
presented by some of the observed molecular tracers,
in Fig.~\ref{fig:cowidth} we show the maps of the $2^{nd}$ moment 
of the data. 
%

%
%

The map of the \COI\ line width in Fig.~\ref{fig:cowidth} clearly shows
that the FWHM in the SE clump is enhanced compared to the rest
of the ambient cloud. Although to a lesser extent, this is also seen
in the NW clump.
The \COII\ line widths are also larger in the SE
clump, which is also an effect of  the presence of the two velocity
components (clumps A, B and the Spur) in this part of the cloud. 
A two-component Gaussian fit in the SE region shows that most of the
contribution to the line width of  \COII(1--0) comes from the lowest
velocity component, although the peak of the line-FWHM is shifted 
relatively to the nominal positions of the clumps A and B.

Likewise, the \N2H(1--0) map shows an enhacement of the line width in
the SE clump, whereas this is less evident in the NW clump, also due
to the limited extension of the map in that region.
The CS(2--1) and \CSI(2--1) maps of the $2^{nd}$ moment of the data shown
in Fig.~\ref{fig:cowidth} show some enhacement of the line width at the
positions of the submillimeter sources, but the most noticeable feature is the
much larger line width at the position of the S68NW starless core 
described by Williams \& Myers~(\cite{WM99}).



\subsection{Physical parameters}
\label{sec:phpar}

In this section we derive the physical parameters of the molecular gas,
using various tracers and approximations. We first derive estimates
of the gas excitation and column densities using a simplified LTE approximation,
then we derive particle densities from the \CSI\ line intensities using
an LVG approximation combining our data with the \CSI(5-4) observations of 
McMullin et al. (\cite{mcm94}). Finally, we provide various estimates of 
the masses of the entire cloud and of the various clumps identified by 
TSOO. 

\subsubsection{Excitation and column densities}
\label{sec:exc}

\begin{figure}
%
%
%
%
%
\centerline{\psfig{figure=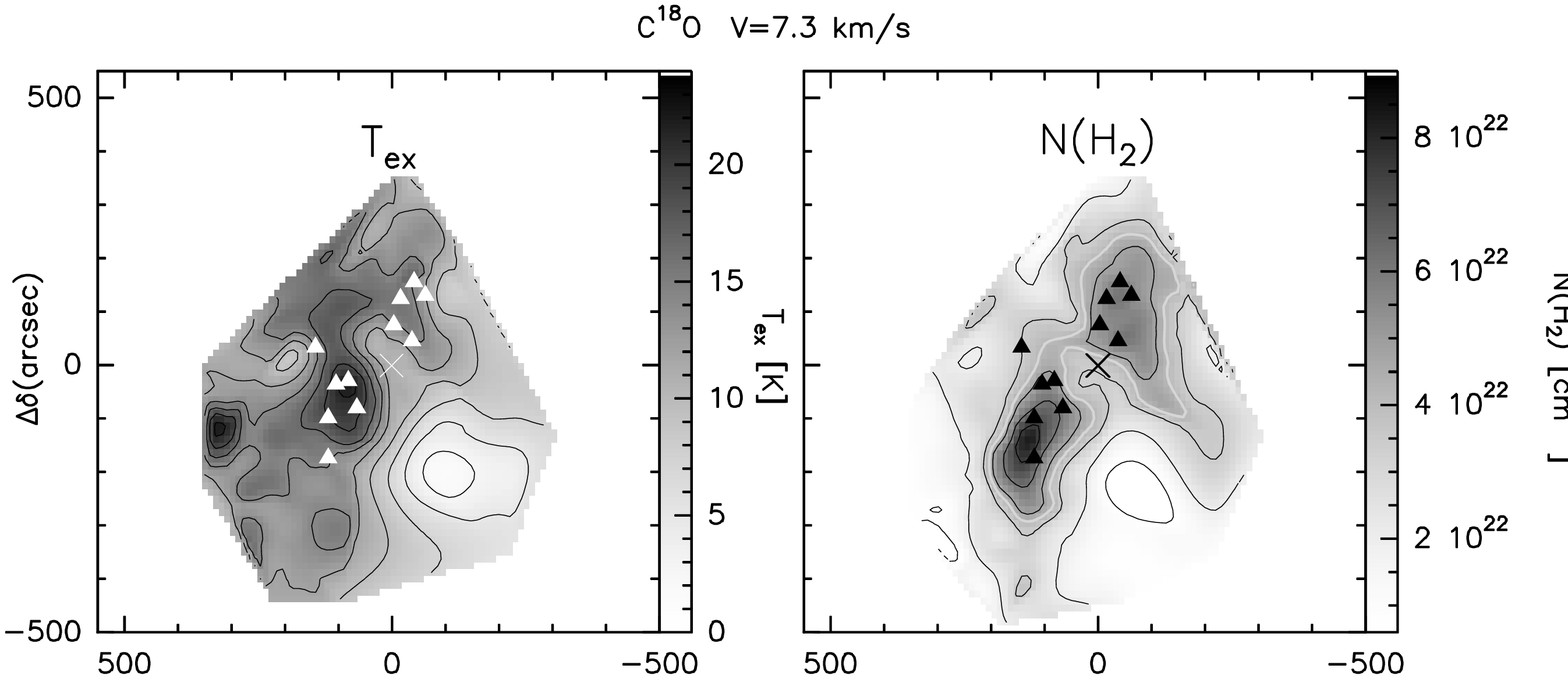,width=8.8cm,angle=270}}
\centerline{\psfig{figure=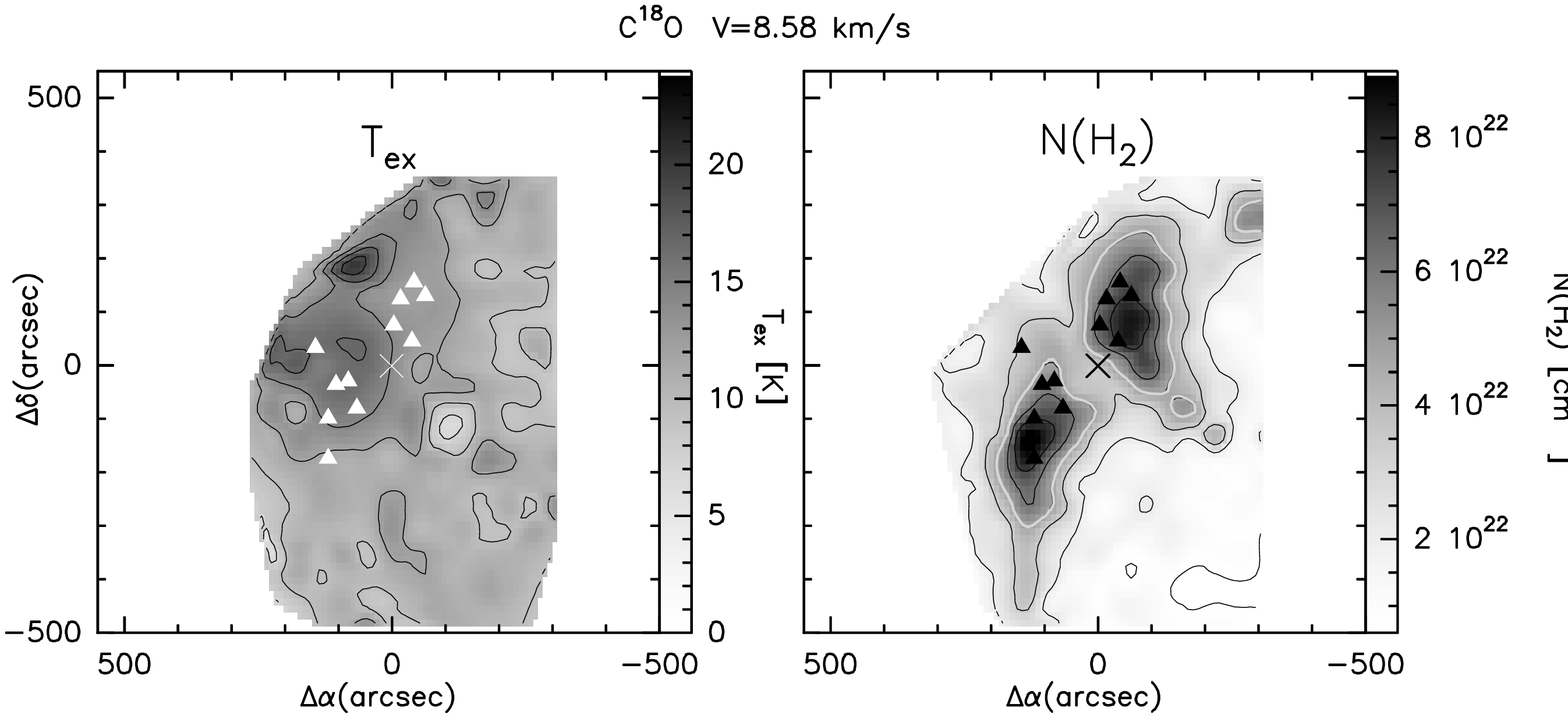,width=8.8cm,angle=270}}
\vspace*{1.5cm}
\caption[ ]{
Maps of $\tex$ and of \HM\ column density using \COII\ and a velocity
of 7.3~\kms (top) and 8.58~\kms (bottom), characteristic of the SE
and NW clumps, respectively.
}
\label{fig:dcolCO}
\end{figure}

The \COII\ emission can be used to trace the column density of the Serpens
cloud. To derive values for the column density we first estimated 
the excitation temperature throughout the entire region, based on the \COI\ line
temperatures at a velocity of 8.58~\kms. We also estimated the \COII\ optical 
depths using the \COI\ to \COII\ line ratio and a relative abundance 
$X$[\COI]/$X$[\COII]$=12$. 
We then used both $\tex$ and $\tau$(\COII) to calculate the column density 
of a $J\rightarrow J-1$ rotational transition according to the following 
formula (Lis \& Goldsmith~\cite{lg91}):
%
\begin{equation}
N_{\rm tot} = \frac{4.0\times 10^{12}}
{J^2\mu^2\,B}
Z \exp\left( \frac{E_{\rm J}}{T_{\rm ex}} \right ) 
\frac{1}{\eta_{\rm mb}} \frac{\tau}{1-e^{-\tau}}
\int \, T_{\rm A}^{\star} \, {\rm d}v
\end{equation}
%
where $B$ denotes the rotational constant, $E_{\rm J}$ is the 
upper state energy, $\mu$ is the dipole  moment and we use the
escape probability $\tau/[1-\exp(-\tau)]$ to account for first 
order optical depth effects. 
We also assume a unity
beam-filling factor. 
The excitation temperature is obtained by inverting the 
equation of transport:
\beq
\tmb = [J_{\nu}(\tex)-J_{\nu}(\tbg)]\,[1-\exp(-\tau)]
\eeq
where $\tex$ and $\tau$ are calculated  at a specific 
velocity and $\tau$ is evaluated as described above.
Moreover, for $k T_{\rm ex} >> h B$ the partition 
function of a linear molecule is given by
\beq
Z = \frac{k T_{\rm ex}}{h B}
\eeq
where $h$ is the Planck constant. 
The distribution of $N_{\rm tot}$(\HM) calculated using \COII\ data is shown in 
Fig.~\ref{fig:dcolCO} and corresponds very well to the map of the \COII\ 
integrated emission of Fig.~\ref{fig:comap}.

The CS and \CSI\ data can be used to derive the gas total column density 
following the method described above for \COI\ and \COII. \CSI\ is in fact
optically thin and can thus be successfully used to trace the column density 
of the gas if its relative abundance can be determined. We derived the 
\CSI\ abundance using the LVG approximation to model the measured line 
intensities, as described later in this section.
Therefore, we can use \COII\ and \CSI\ to
compare the spatial distributions of the column density and excitation
temperature at two velocities characterizing the NW (8.58~\kms) and SE
(7.3~\kms) sub-clumps. We find that CS and \COII\ show a different behaviour:
in fact, the \COII\ maps present a spatial offset between the peaks 
of $\tex$  and those of the gas column density distributions 
(see Fig.~\ref{fig:dcolCO}).
At the velocity of the NW sub-clump the \COII\ column density map shows 
a clump that surrounds the NW group of submillimeter sources, whereas the 
map of $\tex$ shows a smooth plateau at  the same position. At the velocity 
characterizing the SE sub-clump these differences are less extreme:
a broad peak of column density surrounds sources SMM2/4/11, whereas the $\tex$
map peaks between SMM4 and SMM3/6. In both the NW and SE sub-clumps the 
distribution of the column density (see Fig.~\ref{fig:dcolCO}) follows quite 
closely the map of the \COII\ integrated emission shown in the bottom panel of
Fig.~\ref{fig:comap}, suggesting that the emission peaks
are column density maxima and are not due to peaks in the excitation of
the molecular gas.

\begin{figure}
%
%
\centerline{\psfig{figure=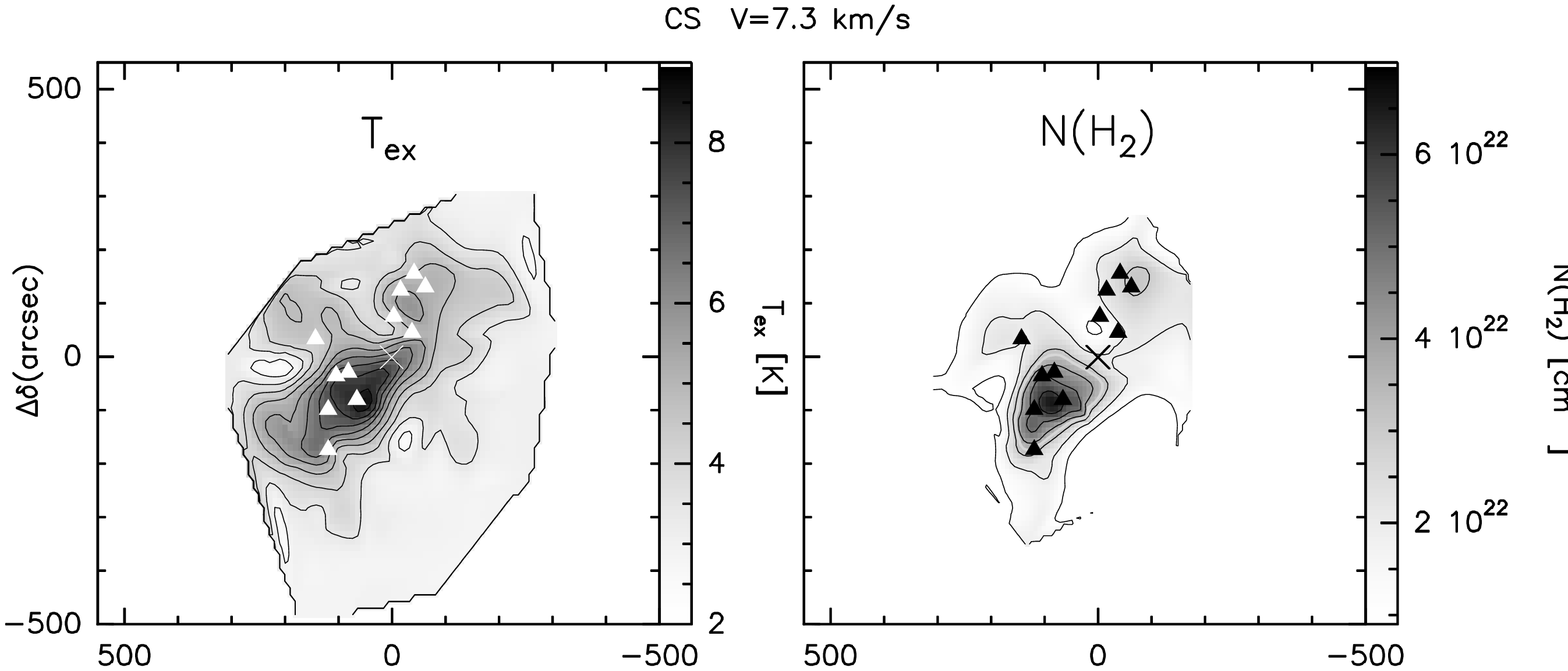,width=8.8cm,angle=270}}
\centerline{\psfig{figure=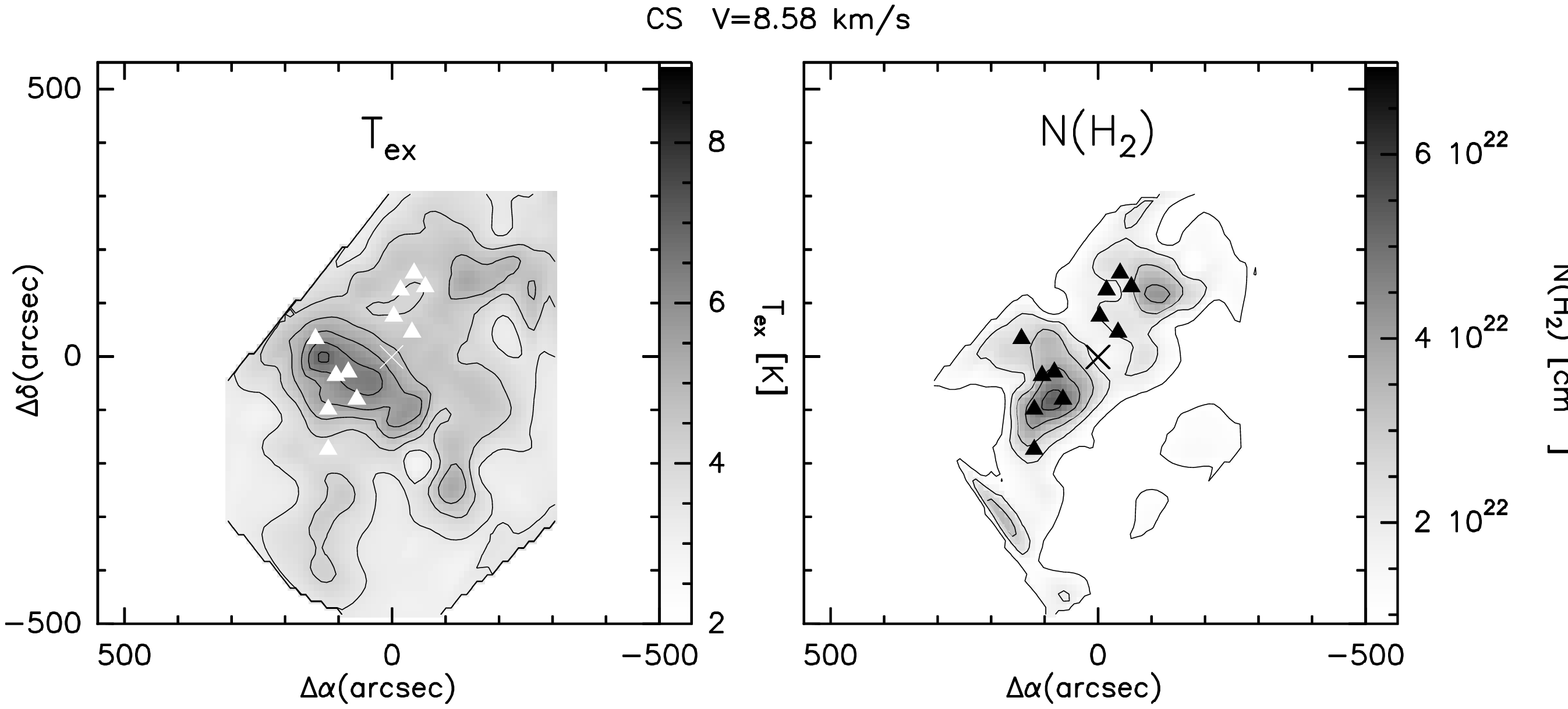,width=8.8cm,angle=270}}
\vspace*{1.5cm}
\caption[ ]{
Maps of $\tex$ and of \HM\ column density using CS and a velocity
of 7.3~\kms (top) and 8.58~\kms (bottom), characteristic of the SE
and NW clumps, respectively.
}
\label{fig:dcolCS}
\end{figure}

The CS and \CSI\ transitions can also be used to determine the spatial 
distribution of $\tex$(\CS) and the column density. The map of $\tex$(\CS)
has a broad peak including the SMM 3/4/6/8 group. 
The gas traced by CS, however, is characterized by spatial distributions of
$\tex$ and column density that are much more similar 
(see Fig.~\ref{fig:dcolCS}). In the NW sub-cluster
the column density map shows a peak at the position of S68NW, while $\tex$ 
has a maximum slightly NW of the S68NW position. At the velocity characterizing 
the SE sub-cluster,  the $\tex$ and column density maps both have a peak very 
close to the position of SMM4 and show little differences otherwise.
Therefore, there is a tendency of the gas column density to have a different
spatial distribution from that of $\tex$  in the NW sub-cluster, and this 
difference is much more evident in \COII. 


\subsubsection{Density estimates}
\label{sec:dens}

The physical number densities towards the densest condensations can be 
derived using density tracers such as the optically thin  \CSI\
transitions. Because we observed the $J=2-1$ line only, we also used the \CSI(2--1) 
and (5--4) observations towards SMM1 of McMullin et al.~(\cite{mcm94}).
By using the $J=2-1$ measurements taken with the FCRAO and NRAO telescopes
we were able to estimate the deconvolved angular diameter of the source 
$\theta_{\rm s}\simeq 70''$. Assuming a gas temperature of 35~K towards SMM1
and a velocity gradient of about 21~\kp, we then used the LVG approximation to 
model the measured line intensities, obtaining a number density 
$\nmh\simeq 5.6\times 10^5$~\cmc\ and a relative abundance
$X$[\CSI]/$X$[\HM]$=4\times10^{-11}$, which are consistent with the values 
obtained by McMullin et al.~(\cite{mcm00}).
There are no \CSI(5--4) observations towards the SE sub-clump. However, an
estimate of the physical conditions at the SMM4 position can
be derived by using the LVG model parameters obtained for SMM1, and changing
one parameter only, e.g. the density, to fit the \CSI(2--1) line intensity.
By doing this we derive $\nmh\simeq 7\times 10^5$~\cmc\ towards SMM4.
%

\subsubsection{Mass estimates}
\label{sec:mass}

Integrating over an area of 111~arcmin$^2$, or 0.90~pc$^2$, the total mass 
of the \COII\ emitting region is about 200-500~\solmass\ depending on the
adopted value for the excitation temperature and assuming a distance to 
Serpens of 310~pc and a \COII\ fractional abundance of 
$X$[\COII]/$X$[\HM]$=10^{-7}$ (McMullin et al. \cite{mcm00}). The 
\COI\ excitation temperature used to calculate the total mass in the
Serpens cloud can be estimated as described in Sect.~\ref{sec:exc},
obtaining values
in the range $\simeq 7$ to 22~K, used to calculate the column densities
in Fig.~\ref{fig:dcolCO}. However, these temperatures are lower than the gas
temperatures derived by McMullin et al.~(\cite{mcm00}), due to the 
fact that the peak \COI\ line temperature traces the cooler larger 
scale cloud.  The range in the
cloud total mass given above reflects these
uncertainties in the value of the excitation temperature.


The maps shown in Figs.~\ref{fig:dcolCO} and \ref{fig:dcolCS} 
are likely to underestimate the column density when the \COI\ excitation 
temperature is used. Therefore, we have estimated the column density using
an overall gas temperature 
of 25~K 
from McMullin et al.~(\cite{mcm00}), it is then
possible to obtain the masses of the molecular gas in the NW and SE 
clumps using the formula:
\beq
M_{\rm cd} = d^2\, m({\rm H_2}) \,\mu_{\rm g} 
\int \, N({\rm H_2}) \, {\rm d}\Omega
\eeq
where $\int \, N({\rm H_2}) \, {\rm d}\Omega$ is the hydrogen column
density integrated over the region enclosed by the contour 
level at 50\% of the peak value for each sub-clump, 
$\mu_{\rm g}$ is the mean molecular weight (1.36), $m({\rm H_2})$
is the mass of a hydrogen molecule and $d$ is the distance to the Serpens
cloud. The masses derived in this way (see Table~\ref{tab:masses})
can  be compared with the virial masses, 
\beq
M_{\rm vr}(M_\odot) = 0.509 \, 
d({\rm kpc}) \, \theta_{\rm s}({\rm arcsec}) \, \Delta v^2({\rm km\,\,s^{-1}})
\eeq
where $\Delta v$
is the average full width at half maximum of the line 
($\sim$1.3 and $\sim$1.6~\kms for \COII\ and \CSI, respectively). The typical
diameter of the source is obtained as $\theta_{\rm s}=\sqrt{\Omega_{\rm s}}$,
where $\Omega_{\rm s}$ is the solid angle, within the 50\% contour,
 subtended by each sub-clump 
and measured using Figs.~\ref{fig:dcolCO} and \ref{fig:dcolCS}.
The masses of the two condensations can also be constrained assuming
a spherical geometry and the peak density derived from a statistical 
equilibrium  model applied to the \CSI\ transitions (see above), and
using
\beq
M_{\rm se}(M_\odot) = 2.926 \times 10^{-9} \, d^3({\rm kpc}) \, 
\theta^3_{\rm s}({\rm arcsec}) \, \nmh ({\rm cm^{-3}})
\eeq
Clearly, $M_{\rm se}$ should 
be considered an upper limit to the mass of the molecular gas as we have 
used the peak and not average density values, and it is also much more 
sensitive to the error on the distance to the cloud. 
In Table~\ref{tab:masses} we present a summary of the various mass 
determinations for the two sub-clumps SE and NW, including the values 
determined in TSOO and TS98 for the N$_2$H$^+$(1--0) virial mass and for 
the total mass of the dust cores and stellar components within the clumps.

%
%
\begin{table*}
\begin{flushleft}
\caption[ ]{
$M_{\rm cd}$ is the mass obtained from the column density integrated
over an area corresponding to 50\% of the peak value. $M_{\rm vir}$ is the
virial mass, also obtained using the diameter of the 50\% contour level, and
$M_{\rm se}$ is the mass of a spherical clump with uniform density equal
to the peak hydrogen density determined from statistical equilibrium models.
M$_{\rm d}$ is the total mass of compact dust cores and M$_\ast$ is the 
estimated total mass of near infrared stars.
}
\begin{tabular}{lccccccccc}
\hline
Sub-clump   & $M_{\rm cd}^{{\rm (a)}}$
            & $M_{\rm vir}^{{\rm (a)}}$
            & $M_{\rm cd}^{{\rm (b)}}$
            & $M_{\rm vir}^{{\rm (b)}}$
            & $M_{\rm se}^{{\rm (b)}}$
            & $\theta_{\rm s}^{{\rm (b)}}$
            & $M_{\rm vir}^{{\rm (c)}}$
            & $M_{\rm d}^{{\rm (c)}}$
            & $M_\ast^{{\rm (c)}}$\\
            & (M$_\odot$)              & (M$_\odot$) &(M$_\odot$)&(M$_\odot$)
            & (M$_\odot$)             & (arcsec)   &(M$_\odot$)&(M$_\odot$)\\
\hline
NW     & 60   & 38 & 54   & 58   & 130 & 139 & 27 & 19.5 &4  \\
SE     & 50   & 41 & 80  & 65   & 220 & 154 & $\sim$30 &$\sim$12 & $\sim$7   \\
\hline
\end{tabular}
\label{tab:masses}

\vspace*{1mm}
$^{{\rm (a)}}$ Assume $\tex=\tk=25$~K (McMullin et al.~\cite{mcm00}), and $\Delta v\sim$1.3~km\,s$^{-1}$ \\
$^{{\rm (b)}}$ Assume $\tex=\tk=30$~K (McMullin et al.~\cite{mcm00}), and $\Delta v\sim$1.6~km\,s$^{-1}$  \\
$^{{\rm (c)}}$ Values from TS98, TSOO, and references therein\\
\end{flushleft}

\end{table*}

\section{Discussion}
\label{sec:disc}

\subsection{Large-scale cloud kinematics}
\label{sec:rot}

%
%
%
%
%

As discussed in Sect.~\ref{sec:kin}, our \COI\ and \COII\ data show a 
large-scale gradient in the line centroid velocity, which appears to be 
consistent with what expected if the cloud is in slow rotation.
%
%
%
Using the \COII\ value for the velocity gradient found in 
Sect.~\ref{sec:velfit} we find an angular velocity 
$\omega \simeq 4.1 \times 10^{-14}$~s$^{-1}$ and
following Goodman et al. (\cite{good93}) we can then calculate the 
parameter $\beta$, defined as the ratio of rotational kinetic energy to
gravitational energy. For a sphere with constant density, $\rho_{\circ}$,
$\beta$ can be written as:
\beq
\beta = \frac{1}{4 \pi G \rho_{\circ}} \, \omega^2
\label{eq:beta}
\eeq
wehere $G$ is the gravitational constant. To calculate $\rho_{\circ}$ we 
use the total gas mass, $\sim$300~\solmass\  
(corresponding to an 
extended region that includes  both the NW and SE clumps and the lower 
density envelope) 
and the corresponding average radius, 0.55~pc. Hence, 
$\rho_{\circ}\simeq 2.9 \times 10^{-20}$~g~cm$^{-3}$.
Substituting in Eq.~(\ref{eq:beta}) we find
$\beta \simeq 0.07$. All clouds studied by Goodman et al. (\cite{good93})
and Barranco \& Goodman~(\cite{BG98})
have $\beta < 0.18$, which is low enough to inhibit, according to those
authors, fragmentation due to rotationally driven instabilities. 
Hence, the fragmentation observed in Serpens may have involved both
gravitational and magnetic instabilities, but it is very unlikely that
the cloud rotation could have induced fragmentation.
Incidentally,
we note that the direction of the velocity gradient is approximately
perpendicular to the average outflow axis
determined by TSOO (see also Table~\ref{tab:toutflows}).
It is tempting to speculate that the global cloud
rotation may have influenced the orientation of circumstellar disks
surrounding each individual (proto-)cluster member; however, 
at this time this is a pure speculation since we cannot prove whether there
is a physical connection between these two similar directions.
The relationship between the large-scale velocity gradients and the
young stellar objects embedded in the Serpens core should be 
further studied by measuring the velocity gradients in the 
circumstellar disks surrounding most of the objects (see e.g. Hogerheijde
et al.~\cite{Hea99}).

\begin{table}
\begin{flushleft}
\caption[ ]{
Outflows directions position angles from the compilation of TSOO
}
\begin{tabular}{lcrc}
\hline
Source & Outflow direction & Outflow tracer & Reference$^{{\rm (a)}}$ \\
       & [deg E of N]      &                &   \\
\hline
S68N & 140 & CS, CH$_3$OH & 1,2\\
A3   & 140 & H$_2$ & 3\\
SMM5 & 150 & Refl. Nebula & 4\\
SMM1 & 140 & H$_2$, CS, CO  & 1,2,3\\
SMM3 & 170 & H$_2$, CS & 1,5\\
SMM4 & 180 & H$_2$, CS & 1,6,7\\
\hline
Average & $\sim$155 & & \\
\hline
\end{tabular}
\label{tab:toutflows}
\vspace*{0.1cm}

$^{{\rm (a)}}$ 1: TSOO; 2: Wolf-Chase et al.~(\cite{WCea98}); 
3: Hodapp~(\cite{H99}); 4:  Kaas~(\cite{K99}); 
5: Herbst et al.~(\cite{Hea97}); 6: Hogerheijde et al.~(\cite{Hea99}); 
7: Eiroa et al.~(\cite{Eea97}). 

\end{flushleft}
\end{table}

The simple rigid body rotation approximation used to derive $\omega$ and
$\beta$ may be questioned in a cloud core that displays a large non-thermal
contribution to the line width, an indication of complex motion on scales much
smaller than the beam width. Nevertheless, Burkert \& Bodenheimer~(\cite{bb00})
have shown that turbulent cloud cores with random gaussian velocity fields may
have a net angular momentum different from zero and may display a global
velocity gradient. Surprisingly, the values of the specific angular momentum
derived assuming a rigid body rotation represent a good approximation to the
real values induced by the turbulent motions, at least within a factor of a 
few (Burkert \& Bodenheimer~\cite{bb00}).  Our conclusion
that the measured cloud angular momentum is not large enough to play a dominant
role in the fragmentation of the cloud and on the formation of the cluster is
thus robust.

At present we cannot exclude that the observed velocity gradient is produced
by a more complex kinematical structure of the cloud, such as a shear motion
or a non-spherically symmetric contraction of the cloud.
A shear motion would imply a net angular momentum of the cloud, similar to
the rigid rotation discussed above. In fact, a shear motion can be seen
as a complex rotational pattern, and would not change significantly
the discussion above.

Perhaps more interesting would be the possibility of a global, 
non sperically symmetric cloud contraction. A comparison of the total cloud mass as 
derived from the C$^{18}$O column density ($\ga$300~M$_\odot$)
and the virial mass ($\sim$150~M$_\odot$), suggests 
that a global contraction may be
possible. This would also be consistent with the infall signature detected in
the innermost regions of the cloud (see Sect.~\ref{sec:kin}
and~\ref{sec:phpar}). However, the measured velocity gradient (see 
Table~\ref{tab:vfit} and Fig.~\ref{fig:13cocv}) would 
imply peak contraction velocities $\ga 0.5$~km\,s$^{-1}$,
which is larger than the infall velocities measured in the inner regions
of the cloud (see Table~\ref{tab:vin}). 
We thus conclude that this is an unlikely possibility.

\subsection{Cloud contraction and large-scale physical conditions}
\label{sec:phys}

In Sect.~\ref{sec:phpar} we showed that the cloud core displays two 
column density maxima roughly coincident with the two sub-clump 
identified by TSOO. There is also a strong correlation between the 
high column density regions and the infalling signature in the 
normalized centroid velocity difference map (Sect.~\ref{sec:kin} and
Fig.~\ref{fig:diffcv}).
%
%
As noted in the higher resolution N$_2$H$^+$(1--0)/CS(2--1)
study of the NW sub-clump by Williams
\& Myers~(\cite{WM00}), it appears that the infalling speeds are higher in the 
high column density regions surrounding the {\it already formed} young
stellar objects but not at their precise location,
where the feedback produced by protostellar outflows
is acting against the infall. Thus, our observations, together with
those of Williams \& Myers, support the view that 
the conditions for collapse are met in the higher density regions of the 
cloud, which are globally contracting. At a few selected locations the 
collapse has already produced YSOs that, by means of their molecular outflows,
are starting to revert the accretion and disperse the cloud core.

We have also calculated the turbulent crossing time for
the main sub-clumps in Serpens. The crossing time is defined
as the radius divided by the Gaussian dispersion in
internal velocity (see Sect.~\ref{sec:normvel}). We used the 
solid angle subtended by the 50\% contour (of the 
corresponding peak value) to calculate 
the radius of the main \N2H\ cores of emission, B, C and D.
We repeated the same operation for the core E and for the SE
sub-clump mapped in the \CSI(2--1) line. The crossing times
are remarkably similar, varying in the interval 
$1.2 \times 10^5$ to $1.5 \times 10^5$~yr, although the radii vary
by a factor of 2 (0.055~pc to 0.115~pc) and the velocity dispersions 
by almost a factor of three (0.35 to 0.98~\kms).
In the case of the NW sub-clump these values are consistent
with the time needed to reach chemical steady state, i.e.
several times $10^5$ years according to McMullin et al. 
(\cite{mcm00}). These large cores coexist in Serpens with smaller and
younger regions (e.g., Williams \& Myers \cite{WM99}, \cite{WM00}, 
Narayanan et al. \cite{nmwb01}) suggesting a wide range of 
evolutionary stages.


\subsection{Star formation efficiency in Serpens}
\label{sec:sfe}

Combining our data with the results of TS98, TSOO, Giovannetti et
al.~(\cite{Gea98}), Kaas~(\cite{K99}), and White et al.~(\cite{Wea95}) we can
estimate the average and local star formation efficiency in the Serpens cloud
core.  In Table~\ref{tab:masses} we summarize the mass estimates for the
gaseous sub-clumps, for the dust cores and for the young stellar objects
embedded within them. The total mass of the Serpens 
cloud 
has been estimated
to be in the range 300 (see Sections~\ref{sec:rot}, \ref{sec:mass} and McMullin 
et al. \cite{mcm00}) to 1500~M$_\odot$ (White et al. \cite{Wea95}), 
whereas the total mass of the young stellar
cluster is estimated to be $\sim$25-40~M$_\odot$ to which one should add the
mass of forming stars in the dust cores of TS98. Assuming that the cores will
produce stars with a 50\% efficiency, meaning that half of each core mass will
end up in a young star and half will be ejected in a protostellar outflow, the
total mass of the final cluster is estimated in the range 40-60~M$_\odot$. The
{\it global} star formation efficiency is thus estimated in the range 3 to
10\%. However, within the two sub-clumps defined by the N$_2$H$^+$(1--0) 
FWHM contours, where the gas column density is higher, 
the global infall motion is detected (Sect.~\ref{sec:phys} and Williams
\& Myers~\cite{WM00}) and where most of the submillimeter prestellar cores and 
protostars are located (TS98, TSOO),
the {\it local} star formation efficiency is much higher
(25 to 50\%), suggesting that where the collapse conditions are met the star
formation process proceeds with a high efficiency.

\section{Conclusions}
\label{sec:concl}

We performed wide-field imaging (up to about $16'\times 16'$)
of the emission towards the Serpens cloud core using moderately 
optically thick ($^{13}$CO(1--0) and CS(2--1)) and optically thin 
tracers (C$^{18}$O(1--0), C$^{34}$S(2--1), and N$_2$H$^+$(1--0)). 
Our main goal was to study the large-scale distribution of the molecular 
gas in the Serpens region and to understand its relation with the denser 
gas in the cloud cores, previously studied at high angular resolution.

The \COI\ and \COII(1--0) emission traces the distribution of the column 
density in the Serpens cloud. The \COI\ emission is extended and encompasses 
both the NW and SE clusters of submillimeter continuum sources, whereas the
\COII\ emission shows the clear demarcation between the NW and SE 
sub-clumps, which are also separated in velocity (see also TSOO). 
Maps of the CS and \CSI(2--1) emission show several 
condensations in the two main sub-clumps, NW and SE, but present otherwise a 
different appearance from \N2H. The latter
follows the spatial distribution of the
submillimeter sources in both the NW and SE sub-clumps much more closely
than the other molecular tracers. 
In both the NW and SE sub-clumps the distribution of the column density
follows quite closely the map of the \COII\ and \CSI\ integrated emission,
suggesting that the emission peaks are column density maxima and are not
due to peaks in the excitation of the molecular gas. We also find that the SE
and NW sub-clumps are virialised with masses typically of order $60\, M_\odot$
and peak number densities of order $\nmh\simeq 7\times 10^5$~\cmc.

%
The  \COI\ and \COII(1--0) maps of the centroid velocity show an 
increasing but smooth gradient in velocity from E to W. 
%
Using most of the data in a map at once,
by least-squares fitting maps of line-center velocity for
the direction and magnitude of the best-fit velocity gradient, we
find that \COII\ has a 1.25~\kms~pc$^{-1}$ velocity gradient which
is directed  from E to W.
%
%
The velocity gradient is extended to the whole Serpens cloud and is unlikely
that the \COI\ {\it and} \COII\ centroid velocities are seriously affected by
the outflows.  We can also reasonably exclude that the observed velocity
gradient is due to separate clumps with different systemic velocities,
or that it is due to a global, non-symmetric cloud contraction. We thus
think that the observed velocity gradient may indeed be caused by a global
rotation of the Serpens molecular cloud whose rotation axis is very close 
to the SN direction, similar to the average outflow axis found by TSOO.
Following Goodman et al. (\cite{good93}) we also calculated the
parameter $\beta$, defined as the ratio of rotational kinetic energy to
gravitational energy. We find $\beta \simeq 0.07$ which, according to 
Goodman et al. (\cite{good93}), is low enough to inhibit fragmentation due 
to rotationally driven instabilities.  
The cloud angular momentum is not sufficient for being dynamically important 
in the global evolution of the cluster. Nevertheless, the fact that the outflows
are roughly aligned with it may suggest a link between the large-scale 
angular momentum and the circumstellar disks around individual protostars
in the cluster. However, our data does not allow to prove this connection. 
%

The normalized centroid velocity difference, $\delta V_{\rm c}$, is a 
typical indicator of underlying infall. Using CS and \CSI\ we
find two large regions of the map where the negative values of 
$\delta V_{\rm c}$ are more concentrated, which are approximately 
coincident with the SE and NW sub0clumps. 
We find that the SE and 
NW sub-clumps have greater non-thermal than thermal motions, and we 
calculate infall velocities of about 0.1 to 0.4~\kms. 
Together with the results of Williams \& Myers~(\cite{WM00}), 
our observations suggest that infalling conditions are met in a relatively
large region
where the column density of the 
gas exceeds a certain threshold; when a protostellar
object is formed, its outflow locally acts against the infall.

Whithin the high column density regions, where the infalling signature
is detected, the {\it local} star formation efficiency is much larger
than in the entire cloud. This is consistent with the picture that 
where the conditions for star formation are met, then the process is highly
efficient. Given that outflows from young protostars appear to oppose
locally to the infall, such high star formation efficiency can be achieved
only if most of the objects in the cluster are formed over a very short 
timescale. This short formation timescale, which does not allow much time for
interaction between protostars given the relatively low density of objects
observed in Serpens, and the indication that the prestellar cores mass
function is consistent with the field stars IMF (TS98), suggest that 
dynamical interactions and competitive accretion may not be the
dominating processes in these types of regions.


The observational connection of the large-scale rotation and the 
individual stellar 
systems flows, and the extended slow inflow motion limited to the high column
density regions, appear to be consistent with a slow global evolution
of the cloud that locally reachs the conditions for gravitational collapse
and star formation. The very short timescale for the formation of the 
bulk of the cluster is advocated by both the accelerated star formation
view of Palla \& Stahler~(\cite{PS00}) or the dynamical star formation 
view of Elmegreen~(\cite{E00}).

It is difficult, from our observations, to give a definitive answer to
whether star formation occurs in a slow quasi static fashion up to the
point of instability or whether the entire process is driven by the fast
dissipation of turbulence in clouds. Our data appear to be in qualitative
agreement with the expectation of a slow contraction followed by a rapid
and highly efficient star formation phase in localized high density regions.
Our evidence is not conclusive, but we provide a series of observational 
constraints with which numerical simulation have to be tested against
quantitatively. From the observational point of view, on the one hand the 
possible connection between the large-scale rotation and the individual 
circumstellar regions should be made more precise by means of high angular 
observations, on the other hand comparisons between wide area studies of the
physical and kinematical conditions within giant molecular clouds 
and the distribution and properties of the young (forming) stars within 
them should be extended to other complexes. 

\begin{acknowledgements}
We thank Matthew Bate, Riccardo Cesaroni, Cathie Clarke, Bruce Elmegreen, 
Ralf Klessen, Mordecai Mac~Low, Francesco Palla, Anneila Sargent and
Malcolm Walmsley
for insightful discussions about the various models, numerical simulations,
and the possible observational tests.
We thank Chris Davis and Joseph McMullin for promptly providing their
published data in electronic form.
This work was sponsored in part by the Advance Research Project Agency,
Sensor Technology Office DARPA Order No. C134 Program Code No. 63226E
issued by DARPA/CMO under contract No. MDA972-95-C-0004.
\end{acknowledgements}


\end{document}